\documentclass{aa}  

\usepackage{graphicx}
\usepackage{txfonts}
\usepackage{natbib}
\usepackage[breaklinks=true]{hyperref} 
\hypersetup{colorlinks=true,urlcolor=blue,citecolor=blue,pdfborder= 0 0 0}
\bibpunct{(}{)}{;}{a}{}{,}    

\usepackage{placeins}

\begin{document} 

\title{Substructure in  {redMaPPer} clusters and its impact on X-ray morphology and scaling relations}
\titlerunning{Substructure in redMaPPer clusters} 
   \author{R. Tuomainen
          \inst{1}
          \and
          A. Finoguenov\inst{1}
          \and
          J. Comparat\inst{2}
          \and
        L. Doubrawa\inst{3,1}
          }

   \institute{Department of Physics, University of Helsinki, Gustaf Hällströmin katu 2A, Helsinki, FI-00014, Finland 
\and
Univ. Grenoble Alpes, CNRS, Grenoble INP, LPSC-IN2P3, 53, Avenue des Martyrs, 38000, Grenoble, France         
\and
Departamento de Astronomia, Instituto de Astronomia, Geofísica e Ciências Atmosféricas da USP, Cidade Universitária, 05508-900, São Paulo, SP, Brazil         
             }

   \date{Received December 16 2024; accepted }

 
  \abstract 
{Numerical simulations of hierarchical structure formation predict that galaxy clusters retain significant dark matter substructure, a signature of their ongoing assembly. This substructure is traced by both the spatial distribution of member galaxies and perturbations in the hot intracluster medium. Merging events significantly impact the thermodynamic state of clusters, introducing scatter in observable mass scaling relations and thereby affecting their use as precision cosmological probes.}
{We statistically quantified the prevalence and properties of substructure in optical galaxy clusters and directly investigated its impact on X-ray morphology and scaling relations, leveraging new data from the DECaLS Legacy Survey and the SRG/eROSITA all-sky survey.}
{We applied the hierarchical density-based clustering algorithm HDBSCAN to the {redMaPPer} galaxy cluster catalog to identify and characterize substructure from the probabilistic membership assignments. This provides a refined membership catalog and a classification of each cluster as containing substructure or not. We then cross-matched this sample with the eROSITA X-ray morphology catalog to correlate optical substructure with a comprehensive set of X-ray morphological parameters. Finally, we analyzed the scaling relation between X-ray luminosity and optical richness for clusters with and without substructure.
}
{Substructure is a common feature, present in approximately 40\% of clusters; a quarter of the full sample exhibits a fractional contribution to richness in excess of 35\%. We find a highly significant correlation between optical substructure and disturbed X-ray morphologies, a trend that is strongest for high-mass clusters. The clusters with substructure also drive a stronger redshift evolution in the scatter of the $L_X-\lambda$ relation. At low redshifts ($z < 0.2$), they display a systematically higher X-ray luminosity at fixed richness compared to relaxed systems.
}
{We demonstrate that substructure identification with {redMaPPer} is viable and essential for enhancing the precision of cluster cosmology. We attribute the enhanced effect of mergers on X-ray properties at low redshifts to the increased density contrast of low-redshift cool cores and longer substructure survival times, which are possibly due to the suppression of disruptive mixing by effects such as magnetic draping. At lower cluster richness, a discordance between X-ray morphology and the merging state indicates a growing relative importance of active galactic nucleus feedback in governing X-ray morphology.}


\maketitle

\section{Introduction}

Galaxy clusters are powerful cosmological probes, providing competitive constraints on the growth of structure in the late-time Universe \citep[for a recent review, see][]{ClercFinoguenov}.
The utility of clusters as cosmological probes hinges on accurately relating their observable properties---such as X-ray luminosity, gas mass, and optical richness---to their total mass. This is achieved through scaling relations. However, these relations exhibit significant intrinsic scatter, a substantial fraction of which is attributed to the complex merger history and dynamical state of clusters \citep{Kravtsov2006}. Clusters grow hierarchically through mergers and accretion \citep{Frenk1999}, and these events can disrupt the thermal equilibrium of the intracluster medium (ICM), alter the gravitational potential, and induce transient features in the measured ICM properties \citep{Poole2007, ZuHone2011}.

A critical manifestation of this dynamical activity is the presence of substructure. In the optical regime, substructure reveals itself as infalling groups of galaxies within a cluster's halo \citep{Dressler1984}. In the X-ray regime, it is evidenced by asymmetries, shocks, and cold fronts in the hot gas \citep{Markevitch2002}, as well as direct detections of the intragroup medium of infalling groups \citep{Haines2018}. Identifying and characterizing this substructure is therefore not merely a taxonomic exercise but a crucial step in understanding the assembly history of clusters and in controlling the systematic biases that currently limit the precision of cosmological constraints \citep{Pratt2019}.
This focus on substructure is motivated by a critical challenge in cluster cosmology: current studies are systematics-limited. As shown by \citet{ClercFinoguenov}, cosmological constraints become inconsistent when using X-ray luminosity versus optical richness, indicating a significant bias. \citet{Damsted24} identified that a primary driver of this discrepancy is the correlation between optical richness and the large-scale environment, potentially due to variations in cluster formation time affecting the richness--mass relation.

The advent of large, uniform surveys has revolutionized this field. In the optical, algorithms like {redMaPPer} \citep{Rykoff2014} applied to the Legacy Surveys \citep{Dey2019} have cataloged millions of clusters and their member galaxies. In the X-ray, the eROSITA (extended ROentgen Survey with an Imaging Telescope Array) instrument aboard the Spectrum-Roentgen-Gamma (SRG) mission \citep{Predehl2021} is providing the first all-sky X-ray survey since ROSAT, yielding a vast catalog of clusters with detailed morphological information \citep{Bulbul2024, Sanders2025}. This multiwavelength dataset enables, for the first time, a statistically robust investigation of substructure across a large population of clusters.

We leveraged these new datasets to perform a comprehensive study of substructure in galaxy clusters. We applied the hierarchical density-based clustering algorithm HDBSCAN \citep{Campello2013, McInnes2017} to the {redMaPPer} galaxy cluster catalog to identify substructure based on the spatial distribution of probabilistically assigned member galaxies. We then cross-matched these optical clusters with their X-ray counterparts in the first eROSITA All-Sky Survey (eRASS1; \citealt{Merloni2024}) and analyzed their X-ray morphological properties using the largest set of morphological parameters compiled to date \citep{Sanders2025}. Finally, we investigated the impact of substructure on the scaling relation between X-ray luminosity and optical richness.

The paper is structured as follows: In Sect.~\ref{data} we describe the substructure analysis of the redMaPPer catalog and present the analysis of the matched eROSITA clusters. In Sect.~\ref{res} we present the results of our analysis and discuss their implications. We conclude in Sect.~\ref{conclusion}. Throughout the paper we assume a flat $\Lambda$ cold dark matter cosmology with $H_0=70$ km s$^{-1}$ Mpc$^{-1}$ and $\Omega_m = 0.3$. Uncertainties are quoted for the $68\%$ confidence level.

\section{Data and methods}\label{data}

\subsection{redMaPPer catalog}\label{rm}

The main sample for this study comprises 4,029,875 optical clusters identified in the DESI Legacy Imaging Surveys \citep{Kluge24} at redshifts $z < 1$ using the redMaPPer algorithm \citep{Rykoff2014}. In this catalog, each galaxy is assigned a probabilistic membership, and a cluster's total richness ($\lambda$) is calculated as
\begin{equation}
\label{richness}
\lambda = \frac{S}{1-f} \cdot \sum p_\text{mem},
\end{equation}
where $S$ is a scale factor that corrects for the survey-limiting magnitude, $f$ is the masking fraction, and $\sum p_{\text{mem}}$ is the sum of membership probabilities.
Since partial cluster coverage can bias substructure detection---a primary focus of this study---we imposed quality cuts of $S < 1.3$ and $f < 0.2$ to select only well-surveyed systems. While this reduces the final sample size by 13\%, it ensures the reliability of our subsequent substructure analysis, for which completeness in absolute cluster number is less critical than data quality for the retained objects.

\subsection{HDBSCAN}\label{hdbscan}

We used the hierarchical density-based clustering algorithm HDBSCAN \citep{McInnes2017}, originally developed by \citet{Campello2013}, to identify substructure within the {redMaPPer} cluster catalog. HDBSCAN is run separately for each redMaPPer cluster, and the input data to the algorithm contains only the sky positions of the member galaxies in the projected plane---meaning no color, magnitude, or redshift information is used. The primary parameter affecting the clustering results is the minimum cluster size. We adopted a value of ten galaxies for the minimum subcluster size, a choice informed by the efficiency results of SPIDERS' (SPectroscopic IDentification of eROSITA Sources) follow-up of {redMaPPer} members \citep{Clerc16} and considerations of the {redMaPPer} completeness corrections \citep{Rykoff2014}. The results of \citet{Clerc16} indicate that optical characterization of low-richness systems can be highly unreliable, mainly due to the increased chance of projection effects.

A key feature of HDBSCAN is its treatment of low-density data points (i.e., galaxies in our case) as noise. This effectively cleans our catalog of galaxies in sparsely populated regions that are unlikely to be genuine cluster members. The parameter \emph{min\_samples} determines the conservatism of the clustering, where a higher value results in more points being classified as noise. For our analysis, we set this to the minimum value, \emph{min\_samples} = 1 galaxy. The parameter $\alpha$, which additionally affects the conservatism of the clustering, was set to 0.05. This value works as a distance scaling parameter; thus, increasing $\alpha$ reduces the influence of local density variations, while decreasing it increases the sensitivity to smaller density fluctuations.

HDBSCAN does not natively facilitate the return of single-component clusters, and configuring it to do so can introduce a bias inherent to the cluster extraction method, known as the Excess of Mass algorithm \citep{Muller1991, Campello2013}. The algorithm computes a hierarchy of clusters and selects the most stable ones. Here, the term ``mass'' refers to a mathematical quantity related to the stability of the clusters during the clustering procedure, and it is not linked to any physical mass of the astrophysical systems. To mitigate this bias, we implemented a two-step analysis procedure:
\begin{enumerate}
\item HDBSCAN was first executed with a configuration that prohibits single-component clusters. This results in the identification of either multi-subcluster structures or the classification of all points as noise.
\item For clusters where no substructure is identified in the first step, we reran the algorithm with parameters adjusted to enable the identification of single-component clusters.
\end{enumerate}

Using the same \emph{min\_cluster\_size} = 10 for this second run would be overly conservative. Therefore, for these singleton clusters, we used the original {redMaPPer} richness, $\lambda$, as an adaptive minimum cluster size, justified by their prior validation as genuine clusters. All other clustering parameters remain unchanged.

This two-step procedure ensures a consistent cleaning process across both the substructure-rich and no-substructure samples. Following the clustering, the cluster richness is recalculated according to Eq.~\ref{richness}, with any galaxies designated as noise by HDBSCAN assigned a membership probability $p_\text{mem} = 0$. For comparison, we also present results from a control analysis where single-component clusters are identified in a single step from the outset.

\subsection{Performance verification}\label{sims}
We validated the performance of the HDBSCAN clustering algorithm by applying it to a mock catalog derived from the Uchuu simulations \citep{Comparat2025}. Our mock sample consisted of a cylindrical volume projected along the line of sight, containing 210 clusters and 2524 individual dark matter halos.

To quantify the algorithm's performance, we calculated two metrics for each simulated cluster with at least ten member galaxies: the fraction of galaxies that are correctly grouped and the fraction that are misplaced. With our chosen parameters, HDBSCAN correctly groups 77.1\% of galaxies, on average. Of the remaining 22.9\%, 8.5\% are incorrectly labeled as noise, while 14.4\% are erroneously assigned as members of other clusters.

Our analysis was performed solely in the 2D projected sky plane. The clustering performance could potentially be improved by executing the algorithm in 3D comoving space, an approach employed in spectroscopic studies that combine methods like \textit{mclust} with friends-of-friends \citep{Tempel2017}. However, this requires spectroscopic redshift information and is therefore reserved for future work.

\subsection{X-ray properties}\label{xray}

The same catalog of redMaPPer clusters was used to identify extended X-ray sources as galaxy clusters in eRASS1 (\citealt{Kluge24, Merloni24}), with detailed follow-up studies providing extensive morphological and physical characterizations \citep{Sanders2025}. This established framework enables us to demonstrate the practical utility of our substructure analysis.

The eROSITA telescope, the primary instrument on the SRG mission, is designed to perform an all-sky survey in the X-ray band. A catalog of 12,247 clusters detected by eROSITA is presented in \citet{Bulbul2024}. \citet{Sanders2025} conducted a comprehensive analysis of the morphological properties of these clusters, producing the largest uniformly derived set of X-ray morphological parameters to date. This set includes established parameters from previous studies, such as central density \citep[e.g.,][]{Ghirardini2022}, cuspiness (inner density slope; \citealt{Vikhlinin2007}), concentration \citep[e.g.,][]{Santos2008}, fit-peak offset, power ratios \citep{BuoteTsai1995}, the Gini coefficient \citep{Abraham2003, Lotz2004}, photon asymmetry \citep{Nurgaliev_2013}, centroid shift, and ellipticity \citep{Ghirardini2022}. The study also introduces novel parameters: the slosh parameter ($H$) and the multipole magnitudes $M_1$--$M_4$.

In addition to these individual parameters, \citet{Sanders2025} define two composite disturbance scores:
\begin{itemize}
\item
$D_\mathrm{shape}$: A combined disturbance score based on shape parameters, including ellipticity ($\epsilon$), slosh ($H$), multipole magnitudes $M_1$--$M_4$, and the fit-peak offset ($F$).
\item
$D_\mathrm{Comb}$: This score incorporates the same parameters as $D_\mathrm{shape}$ but also includes the concentration ($c_{\text{80--800}}$).
\end{itemize}

\noindent In both cases, a value of 1 indicates a maximally disturbed morphology. All parameters are summarized in Table~\ref{parameters_description}. As some parameters are sensitive to the definition of the cluster center, values are computed using both the best-fitting centroid and the peak of X-ray emission; the latter are denoted with an asterisk (*).

\subsection{Optical and X-ray counterparts}\label{matching}

We identified optical and X-ray counterparts by cross-matching the {redMaPPer} and eROSITA catalogs, considering all pairs within $R_{500}$ and with a maximum redshift difference of $0.02(1+z)$. A total of 314 X-ray sources are associated with multiple optical clusters, and 81 optical clusters have multiple associated X-ray sources; these multiple associations will be discussed separately in Sect.\ref{mult}. The remaining sample consists of 6421 unique X-ray sources and optical cluster pairs.

As anticipated, not every X-ray cluster has an identified optical counterpart due to differences in sky coverage between the Legacy Surveys and eROSITA. We performed substructure analysis on the successfully matched counterparts using the method outlined in Sect.~\ref{hdbscan}.

\section{Results and their implications}\label{res}

\subsection{Substructure detection in eROSITA clusters}\label{sub}
By configuring HDBSCAN to permit single-component clusters in an initial single-step run, we identified 6120 eROSITA clusters without substructure and only 301 clusters exhibiting between 2 and 7 subclusters. This significant imbalance indicates a strong bias toward classifying systems as single-component, likely due to the algorithm's inherent conservatism when forced to return singleton clusters. The resulting small sample of clusters with substructure is statistically insufficient for meaningful further analysis. Consequently, we excluded the results from this initial, biased clustering from the remainder of this work.

We instead based our analysis on the two-step clustering procedure described previously. This method identified a total of 2159 eROSITA clusters with substructure (containing between 2 and 11 subclusters) and 4262 clusters without substructure.

A number of these clusters were previously identified as merging systems by \citet{Wen2024}, who analyzed a sample of 338,841 clusters with spectroscopic redshifts from a parent catalog of 1.58 million clusters \citep{Wen2024b}. Their method identified overdensities in the stellar mass distribution for clusters up to $z \sim 1.5$, yielding 39,282 partner systems associated with 33,126 main clusters. Within this sample, 184 of our substructure clusters and 200 of our non-substructure clusters are present among their main clusters. Furthermore, 16 of our substructure clusters and 39 without substructure are cataloged as partner systems in their work.

Focusing on massive clusters ($\lambda > 30$), \citet{Wen2024} identified 7845 systems with merging subclusters. This subset includes 352 clusters where we detect substructure and 183 where we do not. From the same massive cluster sample, they also identified 3446 post-collision mergers, which encompass 243 of our substructure clusters and 118 without substructure.

Overall, the level of agreement between our substructure detections and the independent results of \citet{Wen2024} is considerable. The observed differences can be partially attributed to fundamental methodological distinctions, particularly in the initial assignment of cluster membership. The {redMaPPer} algorithm, which underpins our catalog, limits membership assignment to within the estimated virial radius ($R_{200c}$). This approach effectively filters out very early-stage mergers whose signatures may lie predominantly outside this boundary. Therefore, our results specifically characterize the role of substructure within the virial radius, as defined by {redMaPPer}.

As a result of our analysis, we refined the richness estimates for all clusters and introduced a new measurement for the substructure richness, which we designate as the infalling richness, $\lambda_{\text{infalling}}$. We defined the host cluster to be the subcluster returned by HDBSCAN with the highest richness, denoted $\lambda_\text{host}$. We applied the same corrections used for the total cluster richness, $\lambda$, as defined in Sect.~\ref{hdbscan}.

Numerical simulations indicate that the substructure mass function exhibits a consistent shape when normalized by the host halo mass. This property allowed us to assess the completeness of our substructure detection method. By substituting mass with richness and applying the mass-richness relation by \citet{McClintock2019}, we could compare our empirical results against theoretical expectations.

\begin{figure}[t]
\centering
\begin{tabular}{c}
\resizebox{0.45\textwidth}{!}{\includegraphics{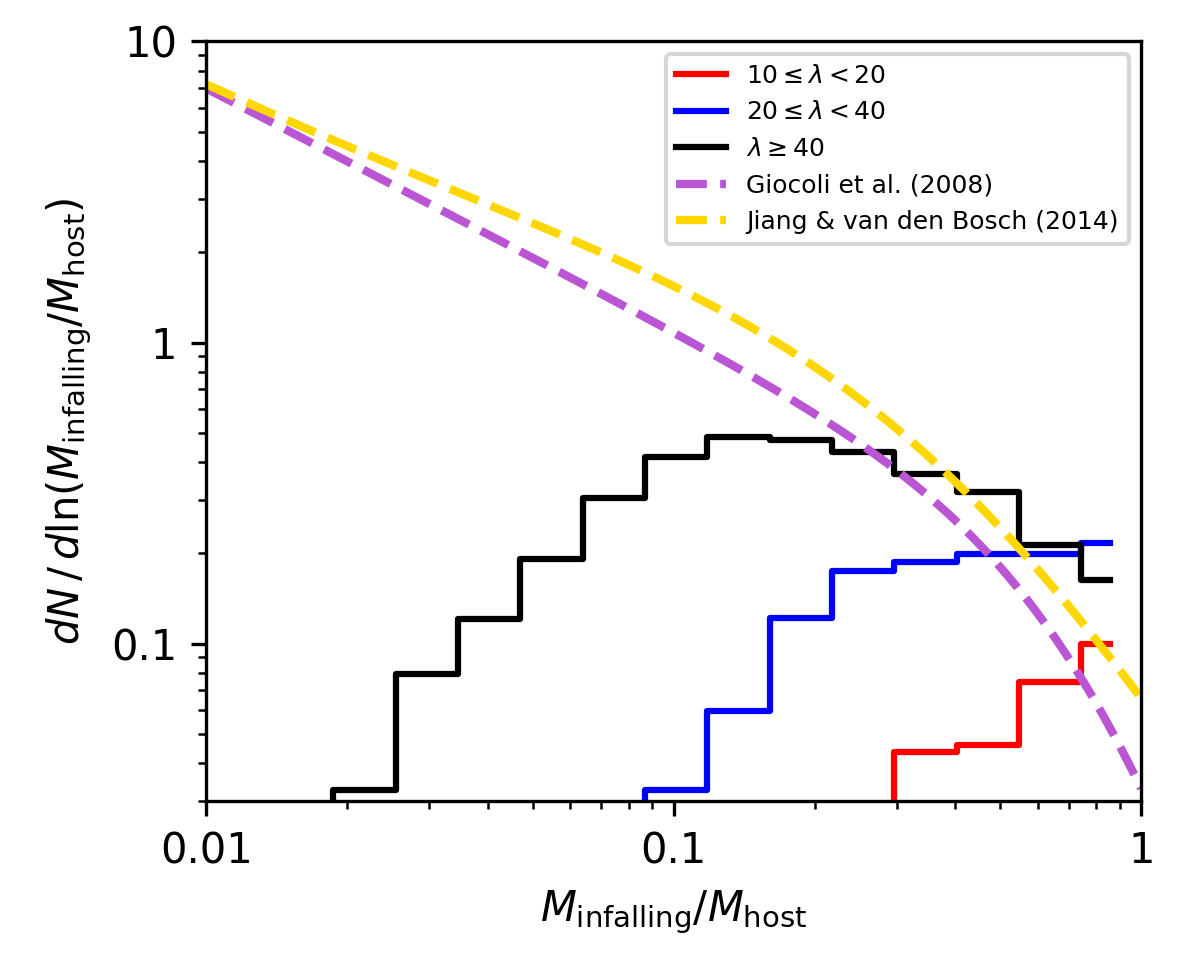}}
\end{tabular}
\caption{Distribution of {mass ratios, $\mu = M_{\text{infalling}} / M_{\text{host}}$}, between the infalling groups and their host clusters {in different bins of total cluster richness. The dashed {purple and yellow} lines show the unevolved subhalo mass functions of \citet{Giocoli2008} and \citet{Jiang2014}, respectively.}
}
\label{richness_ratios}
\end{figure}

Figure~\ref{richness_ratios} presents the distribution of mass ratios between the infalling galaxy groups and their respective host clusters ($\mu = M_{\text{infalling}}/M_{\text{host}}$) for different bins of total cluster richness. We compared our measurements to the unevolved subhalo mass function of \citet{Giocoli2008}:
\begin{equation}
\frac{dN}{d\ln(M_\text{infalling}/M_\text{total})} = N_0 x^{-\alpha}\exp(-6.283x^3), \quad x=\left|{\frac{M_\text{infalling}}{\alpha M_\text{total}}}\right|,
\end{equation}
where $M_\text{infalling}$ is the mass of the infalling subcluster, $M_\text{host}$ is the host mass, $M_\text{total}$ is the total mass of the descendant cluster, $\alpha=0.8$ and $N_0=0.21$. Here, we assume that $M_\text{total}=M_\text{host}+M_\text{infalling}$. Additionally, we plotted the subhalo mass function of \citet{Jiang2014}, which adds an extra power-law term to the fitting function of \citet{Giocoli2008}.

The form and steepness of the mass-ratio distribution of detected infalling subclusters reproduce the unevolved subhalo mass function of \citet{Jiang2014} reasonably well, especially at high richness and high $\mu$ values. At low values of $\mu$, the distribution reflects incompleteness caused by a finite detection threshold. For clusters with $\lambda \geq 20$, our substructure detections are complete down to $\mu > 0.4$, while for $\lambda \geq 40$, completeness extends to $\mu > 0.25$. At the lowest richnesses, only nearly equal-mass mergers ($\mu > 0.8$) are reliably detected.

\begin{figure}[t]
\centering
\begin{tabular}{c}
\resizebox{0.45\textwidth}{!}{\includegraphics{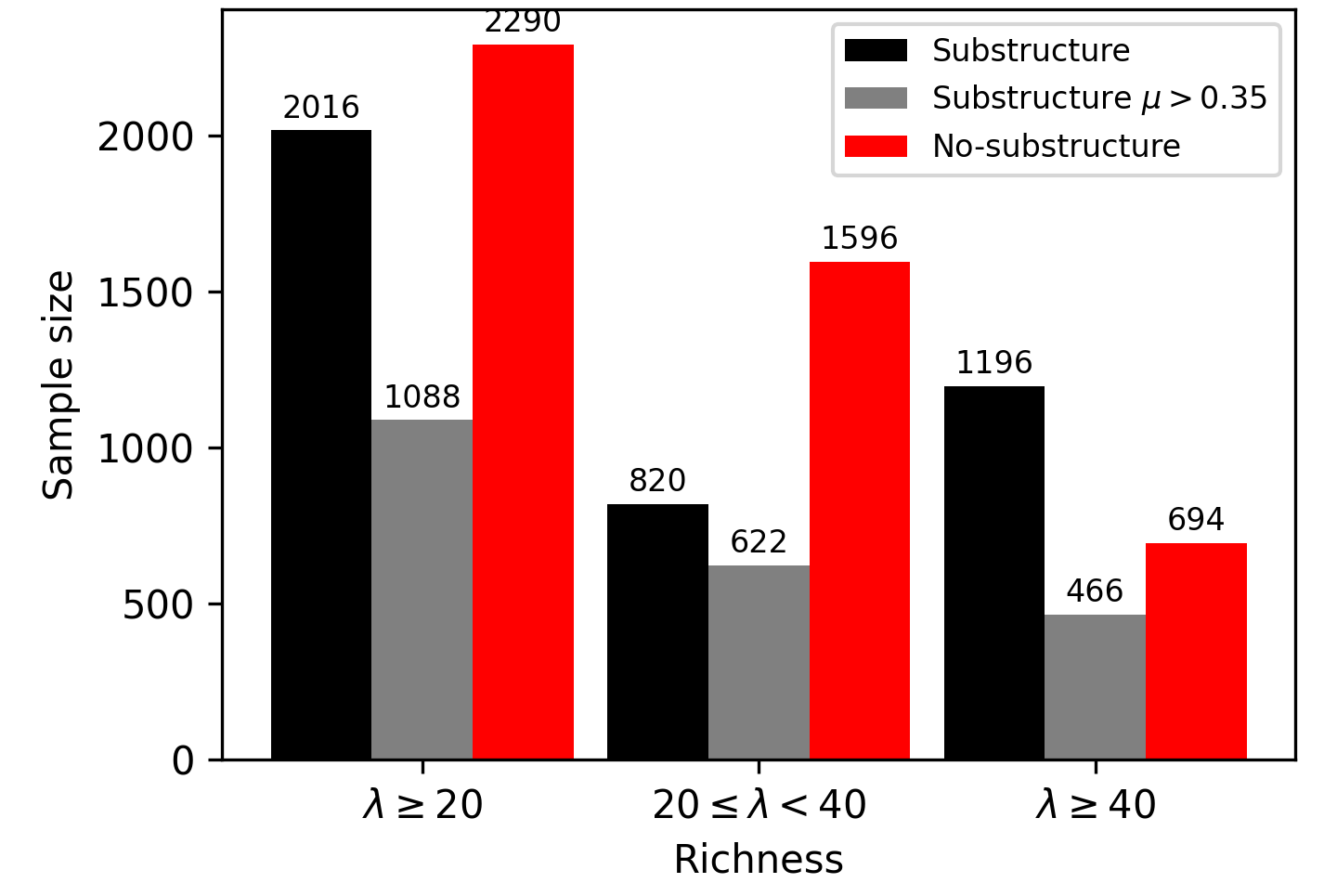}}
\end{tabular}
\caption{Number of substructure and no-substructure clusters for the {full $\lambda \geq 20$ sample and the two richness bins.}}
\label{sample_size}
\end{figure}

The redshift-richness distribution of our sample, shown in Fig.~\ref{redshift_richness}, indicates that incompleteness is primarily a function of low richness, with no strong redshift dependence. Since our substructure measurements are highly incomplete for $\lambda < 20$, we limited the subsequent analysis to $\lambda \geq 20$. We introduced two bins of total cluster richness and additionally created subsamples where we impose a lower limit of $\mu>0.35$, at which the fraction of clusters exhibiting substructure is approximately 25\% in both richness bins. For completeness, we first present the results without the $\mu$ cut and then with the cut applied. Sample sizes for each subset are provided in Fig.~\ref{sample_size}. Single-component clusters are more prevalent at lower richness, while the occurrence of substructure increases with richness due to the finite substructure detection threshold. This trend underscores the importance of the detection threshold: although previous studies report that $40-70\%$ of clusters show substructure \citep{Schuecker2001, Smith2005, WenHan2013}, our results demonstrate that this fraction is highly sensitive to the limits imposed on substructure richness and its fractional contribution to the total cluster richness.

\begin{figure}[t]
\centering
\begin{tabular}{c}
\resizebox{0.45\textwidth}{!}{\includegraphics{figures/figure2.png}}
\end{tabular}
\caption{Redshift vs. richness for the full sample. Clusters with substructure are shown in black, and clusters without in red.}
\label{redshift_richness}
\end{figure}

\subsection{Influence of substructure on X-ray morphology}\label{morph}
We compared the distributions of morphological parameters between clusters with and without identified substructure using the Anderson-Darling (AD) test. Additionally, we used the Kolmogorov-Smirnov (KS) test and a permutation test (shuffle test) with the mean and median as test statistics. The resulting probability values ($p$-values) indicate the likelihood that the two samples are drawn from identical distributions. Since we tested multiple hypotheses on the same dataset, we had an increased likelihood of false positives. This effect can be addressed using a Bonferroni correction; however, the results would be overly conservative since many of the morphological parameters are correlated with each other. We were careful in drawing any conclusions from the results.

In the full $\lambda \geq 20$ sample, without any richness binning or cuts in $\mu$, our analysis reveals statistically significant differences ($p<0.05$) in all morphological parameters between the two groups except slosh ($H$) and multipole magnitudes $M_1, M_3$, and $M_4$. The combined disturbance scores, $D_{\text{comb}}$ and $D_{\text{shape}}$, show that clusters with substructure are generally more disturbed. Their cumulative distribution functions indicate that a larger fraction of these clusters exhibit higher disturbance values at any given threshold. The AD test yields $p$-values of $1.00\times10^{-4}$ and $0.049$ for $D_{\text{comb}}$ and $D_{\text{shape}}$, respectively. The KS test yields a $p$-value of $5.97\times10^{-12}$ for $D_{\text{comb}}$ but fails to find a significant difference between the $D_{\text{shape}}$ distributions, indicating a much stronger distinction in the $D_\text{comb}$ parameter.

\begin{figure}[t]
\centering
\begin{tabular}{c}
\resizebox{0.45\textwidth}{!}{\includegraphics{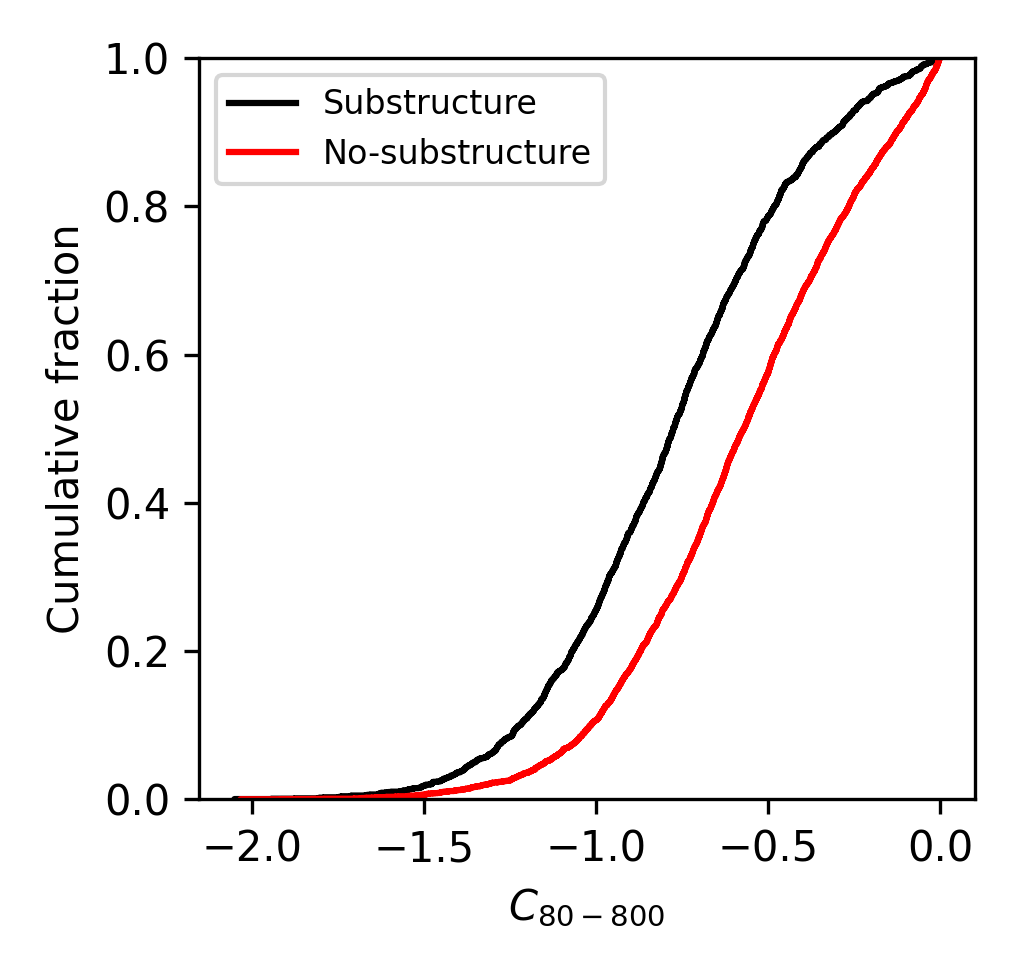}}
\end{tabular}
\caption{Cumulative distributions of the concentration $c_{80-800}$ for clusters {with $\lambda \geq 20.$}}
\label{c80-800}
\end{figure}

This pronounced difference in $D_{\text{comb}}$ is primarily driven by large disparities in the concentration parameter $c_{80-800}$, where clusters with substructure display significantly flatter surface brightness profiles. Concentration is defined as
\begin{equation}
c_{80-800} = \log\frac{I_X(80 \text{ kpc})}{I_X(800 \text{ kpc})},
\end{equation}
where $I_X(r)$ is the integrated surface brightness within radius $r$. A flat profile yields a value near $-2$, while a steep, peaked profile approaches $0$. In order to account for the effects of the point spread function and background, \citet{Sanders2025} compute the concentration using their cluster model, rather than the observed surface brightness. Cluster models are computed with the MultiBand Projector in 2D (MBProj2D) tool \citep{Bulbul2024}. \citet{Sanders2025} use monochromatic 0.2--2.3 keV band images as the input to MBProj2D. The concentration is then evaluated from the model before point spread function convolution and background addition. An AD test on the $c_{80-800}$ distributions returns a $p$-value of $1.00\times10^{-4}$, a KS yields a $p$-value of $3.21\times10^{-17}$, and the permutation test gives $p$-values of $2.0\times10^{-4}$ for both the mean and median, robustly confirming the significance of this result. The distribution of $c_{80-800}$ is shown in Fig.~\ref{c80-800}.

The permutation test does not find significant differences in the means of the photon asymmetry, $A_{\text{phot}}$ and $A_\text{phot}^*$, distributions, while the median shows a clear difference with $p$-values of $1.00\times10^{-4}$ for both distributions. This is caused by the long left tail of the no-substructure sample due to a couple of extreme outliers. Clusters 1eRASS J105803.8-041038 and 1eRASS J030104.7-554410 have $A_{\text{phot}}$ values of $-2935.484$ and $-665.066$, while the median is $\sim0.71$. These outliers increase the variance of the distribution and reduce the robustness of the mean as a test statistic. This showcases the importance of using multiple statistical tests when comparing the distributions. Next, we extended our comparison of morphological parameter distributions to different richness ranges.

In the high-richness bin ($\lambda > 40$), AD tests yield $p$-values of $9.00\times10^{-4}$ for $D_{\text{comb}}$ and $0.016$ for $D_{\text{shape}}$, indicating significant differences between clusters with and without substructure. Beyond the combined disturbance scores, we find significant differences in several other parameters as described in the following paragraph. These include central densities, inner density slope, concentration, fit-peak offset, centroid shift, and the multipole magnitude $M_4$.

As expected for disturbed systems, clusters with substructure also exhibit flatter generalized density profiles, as indicated by the inner density slope parameters $\alpha$ and $\alpha_{50}$, and flatter surface brightness profiles, as measured by the concentration parameter $c_{80-800}$. In contrast, we find no significant difference in the $c_{500}$ distributions, which are computed within a fraction of $R_{500}$ rather than a fixed physical aperture.
Further supporting their disturbed nature, clusters with substructure show higher centroid shift ($\omega$) values, indicating greater spatial variability in the centroid of emission, and higher $M_4$ multipole magnitudes, suggesting more pronounced large-scale asymmetries.
No significant differences are detected in the distributions of several other morphological parameters, including the power ratios $P_{10}$--$P_{30}$, Gini coefficient, photon asymmetry, ellipticity ($\epsilon$), slosh parameter ($H$), and multipole magnitudes ($M_1$--$M_3$). This suggests that the presence of substructure is most evident in a specific subset of morphological indicators.

In the low-richness bin ($20 < \lambda < 40$), differences in shape-based morphological parameters are less pronounced. We find no significant differences in Gini coefficient, ellipticity, slosh, multipole magnitudes $M_1$--$M_4$, or $D_{\text{shape}}$. The only shape parameter showing a significant difference is the fit-peak offset (AD $p$-value = $8.00\times10^{-3}$), with substructure clusters exhibiting larger offsets between the best-fitting center and the X-ray emission peak. Significant differences are also found in non-shape parameters: central densities (lower at fixed $R_{500}$ fraction but higher at fixed 50 kpc radius), inner density slope (flatter for substructure clusters), concentration (flatter profiles in $c_{80-800}$, $c_{80-800}^*$, $c_{500}$, and $c_{500}^*$), power ratios $P_{10}$--$P_{40}$, photon asymmetry, centroid shift, and $D_{\text{comb}}$.

A surprising result in this richness bin is that power ratios $P_{10}$--$P_{40}$ suggest clusters without substructure are less spherical than those with substructure. We note that the difference is minimal, and if we were to use a Bonferroni correction it would not be statistically significant. Therefore, we refrained from drawing any conclusions from this seemingly puzzling result.

\begin{figure}[t]
\centering
\begin{tabular}{c}
\resizebox{0.45\textwidth}{!}{\includegraphics{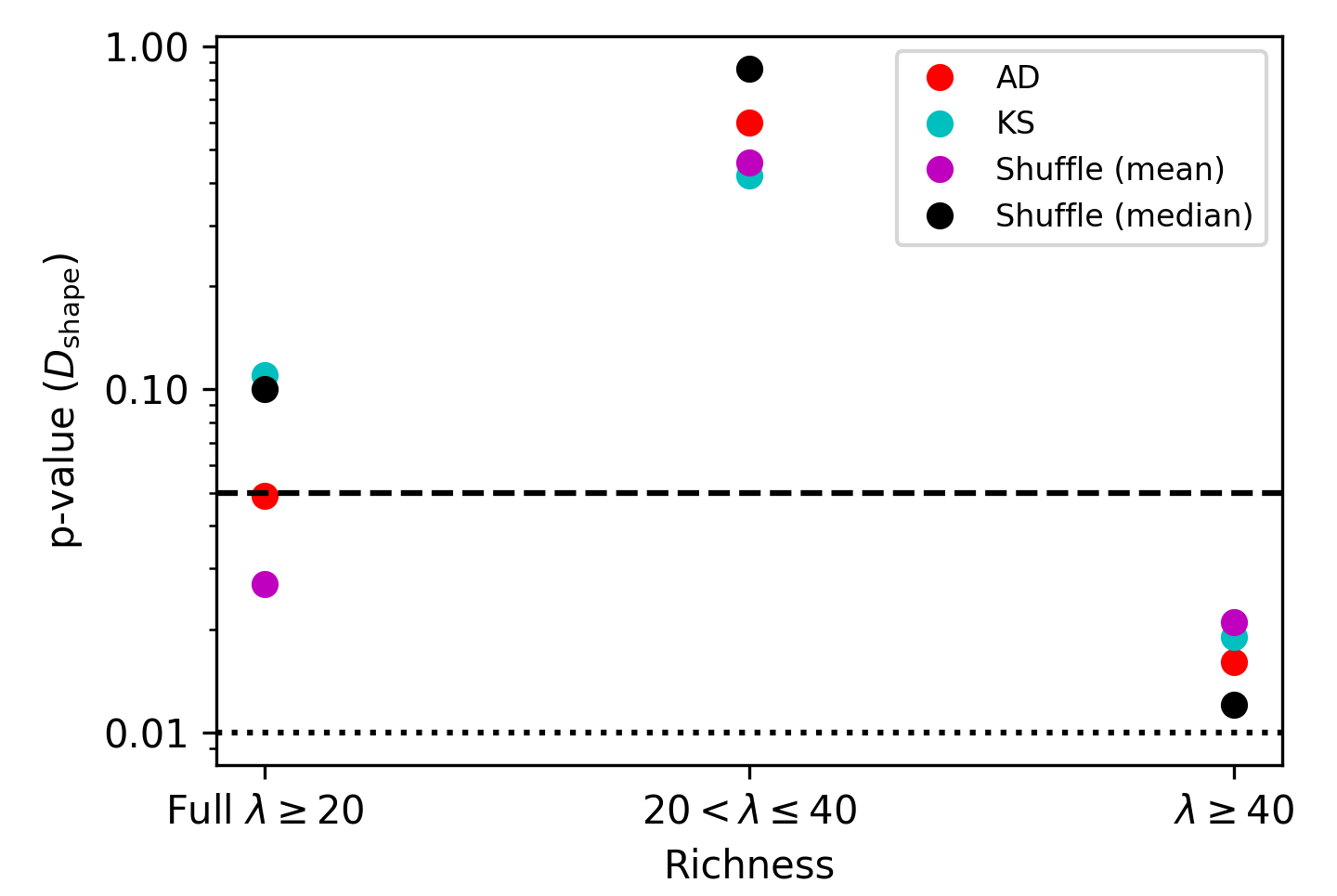}}
\end{tabular}
\caption{$p$-values for the comparison of $D_{\mathrm{Shape}}$ distributions {for the full $\lambda \geq 20$ sample and the two richness bins. Results from the AD tests are shown in red, the KS tests in cyan, and the permutation tests in purple (mean) and black (median).} The horizontal dashed and dotted black lines indicate the $0.05$ and $0.01$ significance levels, respectively.}

\label{richness_vs_p}
\end{figure}

\begin{figure}[t]
\begin{tabular}{c}
\resizebox{\hsize}{!}{\includegraphics{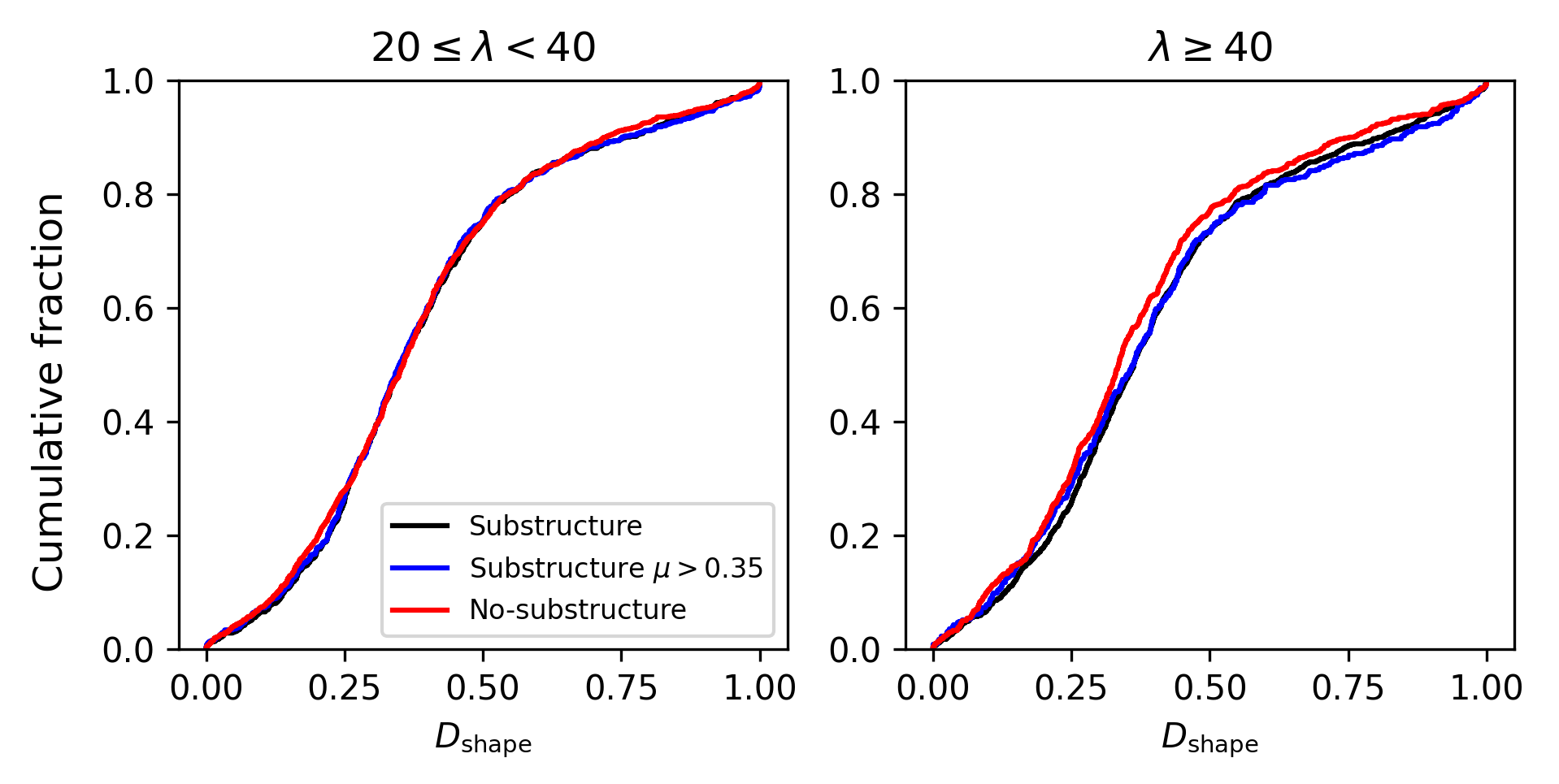}}
\end{tabular}
\caption{{Cumulative distributions of the combined disturbance score $D_{\text{shape}}$ for the substructure (black), substructure $\mu > 0.35$ (blue), and no-substructure (red) groups for two richness bins.}}
\label{d_shape_lambda}
\end{figure}

\begin{figure}[t]
\begin{tabular}{c}
\resizebox{\hsize}{!}{\includegraphics{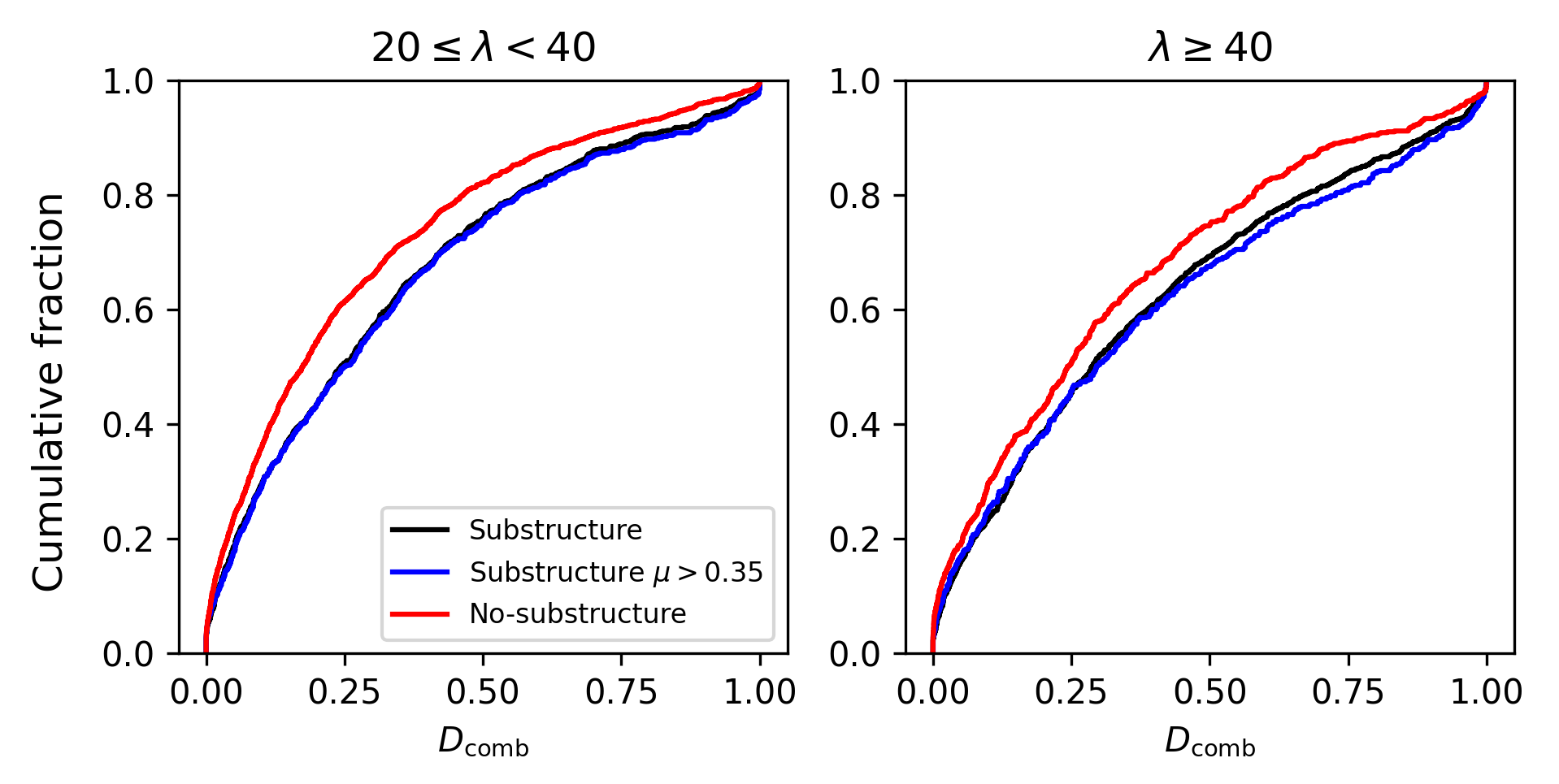}}
\end{tabular}
\caption{{Same as Fig. \ref{d_shape_lambda} but for $D_\text{comb}$.}}
\label{d_comb_lambda}
\end{figure}

The $p$-values for $D_{\text{shape}}$ from AD, KS, and shuffle tests across richness bins are plotted in Fig.~\ref{richness_vs_p}. Additionally, the $D_{\text{shape}}$ and $D_{\text{comb}}$ distributions for the two richness bins are shown in Figs.~\ref{d_shape_lambda} and \ref{d_comb_lambda}. These figures clearly show how the differences in the shape parameters are less pronounced in the low-richness bin. Applying a substructure richness ratio cut ($\mu > 0.35$) does not alter our conclusions regarding disturbance parameters, and we find no significant differences between the full substructure sample and the $\mu > 0.35$ subset, confirming that incompleteness at low $\mu$ does not bias our results. Figures~\ref{d_shape_lambda} and \ref{d_comb_lambda} also include the combined disturbance score distributions for the $\mu > 0.35$ subsamples. These figures demonstrate that the $\mu>0.35$ cut has virtually no effect as the difference is barely noticeable in the plots.

All $p$-values for the statistical tests on the full $\lambda \geq 20$ sample and the two richness bins are summarized in Tables~\ref{p_values_full}, \ref{p_values_l2040}, and \ref{p_values_l40}. Overall, clusters with substructure are consistently more disturbed across all richness levels.

\subsection{Multiple associated sources}\label{mult}

Of the 92 optical clusters with multiple associated X-ray sources, we identify substructure in 48 clusters. Based on visual inspection, the substructure identified by HDBSCAN aligns well with the positions of the eROSITA clusters in 37 of these 48 cases. In the remaining 11 cases, the optically identified substructure does not correspond to the locations of the X-ray sources. Among the 37 clusters where the splitting is successful, we find that 19 contain more subcomponents than the number of associated X-ray sources. We detect no substructure in 44 optical clusters that have multiple X-ray counterparts.

For the 314 X-ray clusters linked to multiple optical clusters, there are 632 optical cluster associations in total. Among these, we find substructure in 315 optical clusters and no substructure in the remaining 317.

\subsection{Scaling relations}\label{scale}
To investigate the impact of substructure on scaling relations, we analyzed the correlation between optical richness and X-ray luminosity. We used the X-ray luminosity in the 0.2--2.3 keV rest-frame band, expressed as $\ln(L_x E(z)^{-1} \, \mathrm{erg^{-1} \, s})$, and the richness as $\ln(\lambda / E_z)$, where $E_z^2 = \Omega_m (1+z)^3 + \Omega_\Lambda$ accounts for the evolution of the scaling relations. The richness was normalized by its median value of 27.9.

We performed parameter estimation using the \textsc{linmix}\footnote{\url{https://github.com/jmeyers314/linmix}} routine from \citet{Kelly2007linmix}, which implements a Bayesian linear regression model that accounts for measurement uncertainties in both variables. The linear relation is given by
\begin{equation}
\eta = \alpha + \beta \xi + \epsilon,
\end{equation}
where $\alpha$ is the intercept, $\beta$ is the slope, and $\epsilon$ represents the intrinsic scatter, assumed to be normally distributed with mean zero and variance $\sigma^2$. The actual values ($\eta, \xi$) are related to the observed values by
\begin{equation}
\begin{split}
x_i & = \xi + x_\text{err} \\
y_i & = \eta + y_\text{err},
\end{split}
\end{equation}
where $x_\text{err}$ and $y_\text{err}$ are the errors in the data $x_i$ and $y_i$.
We ran 100,000 Markov chain Monte Carlo iterations, discarding the first 1,000 steps as burn-in. Reported values correspond to the medians of the posterior distributions.

Results from the \textsc{linmix} fits are summarized in Tables~\ref{tab:scaling_relations} and~\ref{tab:scaling_summary}. We fit the relation for the full sample, as well as separately for clusters with and without substructure.

Our sample covers a wide range in richness and redshift. To ensure high completeness and minimize contamination, we limited our analysis to the region where the eRASS1 sensitivity is high, characterized by a redshift-dependent richness completeness limit $\lambda > z \times 100$, analogous to previous ROSAT All-Sky Survey studies \citep{klein19, Finoguenov20}. Consequently, low-richness clusters are only well sampled at low redshifts. To disentangle redshift and richness trends, we analyzed the scaling relations in bins of both richness and redshift.

\begin{figure}
\centering
\begin{tabular}{c}
\resizebox{0.45\textwidth}{!}{\includegraphics{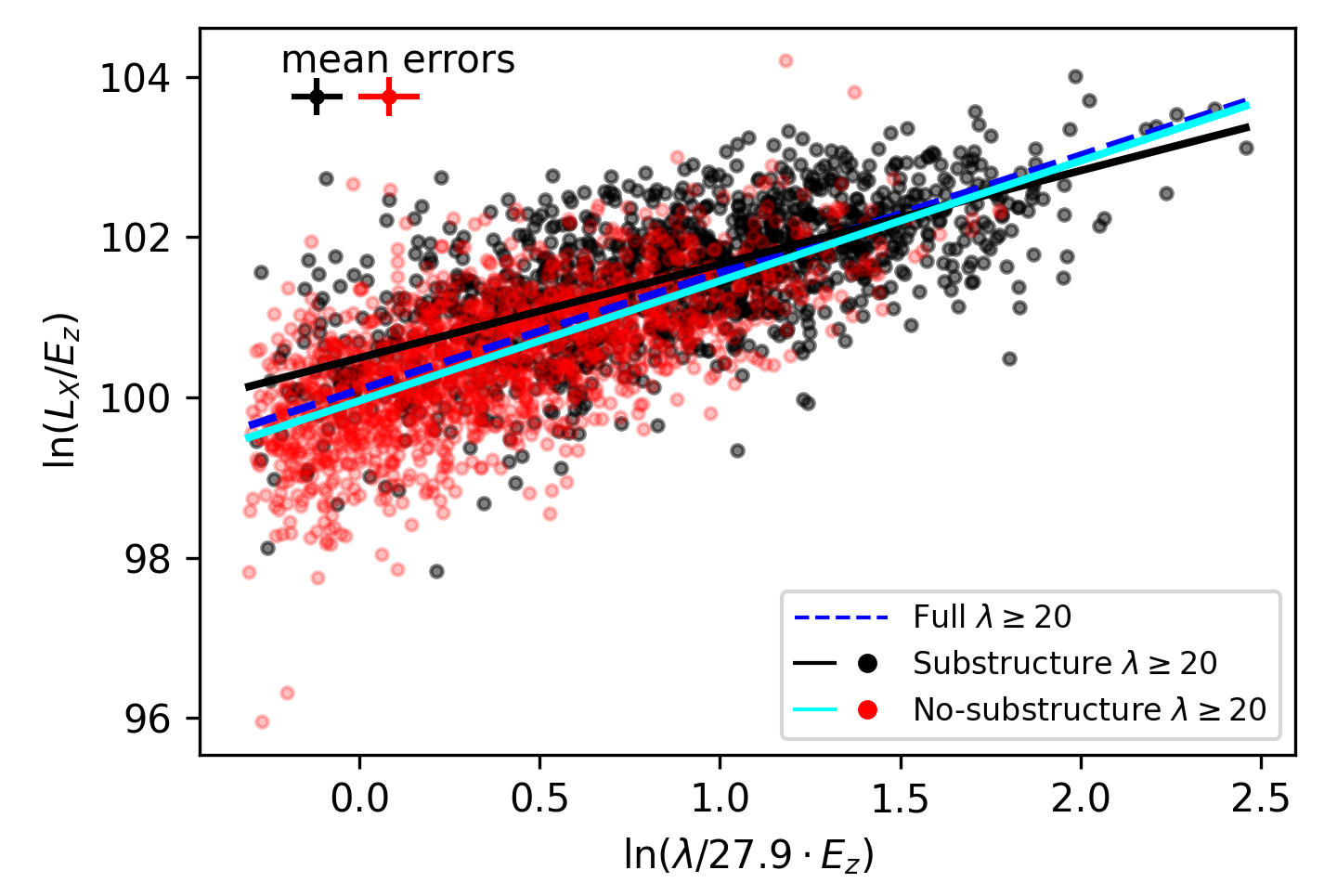}}
\end{tabular}
\caption{X-ray luminosity ($L_X$) vs. richness ($\lambda$) for clusters with $\lambda \geq 20$. Clusters with substructure are marked with black circles, and clusters without as red {circles}. The solid black (solid cyan) line represents the best-fit scaling relation for clusters with (without) substructure. The dashed blue line shows the relation fitted to the full sample ($\lambda \geq 20$). All fits were obtained using the  \textsc{linmix} Bayesian regression method, which accounts for measurement errors in both variables.}

\label{lx_lambda_full_l20}
\end{figure}

\begin{figure}
\centering
\begin{tabular}{c}
\resizebox{0.45\textwidth}{!}{\includegraphics{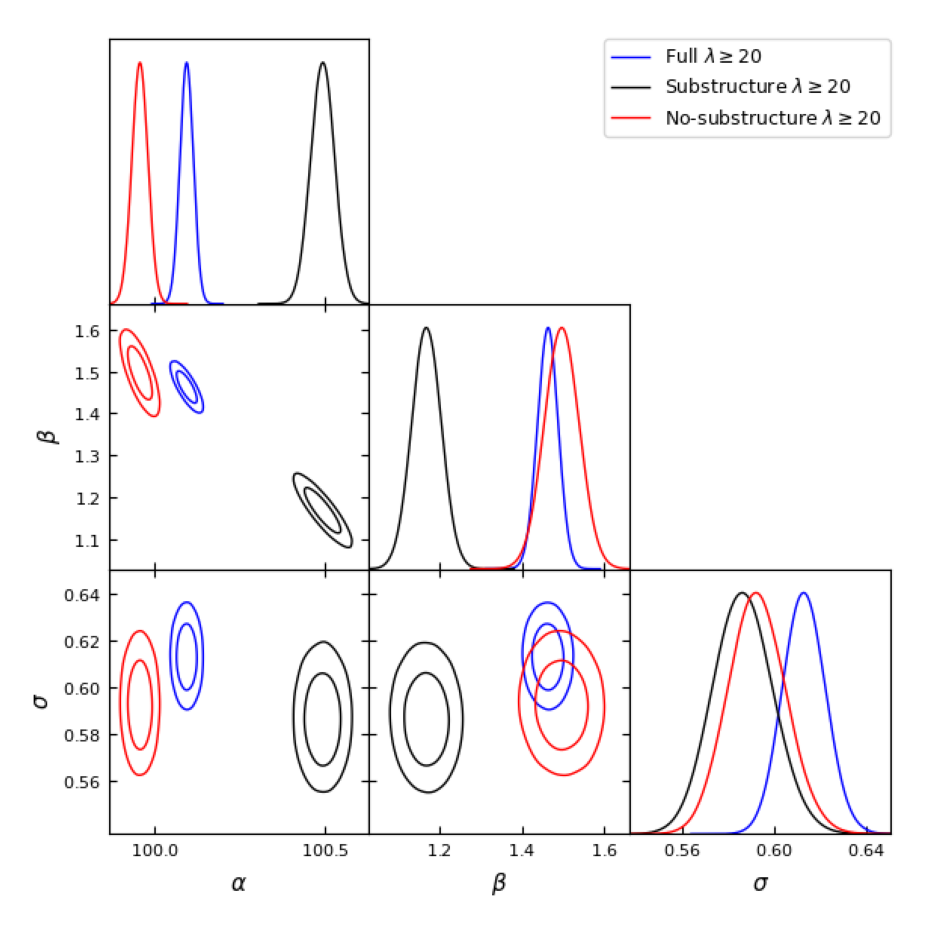}}
\end{tabular}
\caption{Effect of substructure on the X-ray luminosity--richness scaling relation. Contours indicate the 68\% and 95\% confidence regions of the posterior distributions for the scaling relation parameters. Clusters with substructure are represented in black, and clusters without in red. The blue contours and line correspond to the full sample. The parameters $\alpha$, $\beta$, and $\sigma$ denote the intercept, slope, and intrinsic scatter of the relation $\ln(L_X) = \alpha + \beta \ln(\lambda) + \epsilon$, where $\epsilon \sim \mathcal{N}(0, \sigma^2)$.}
\label{triangle_l20}
\end{figure}

Given that substructure measurements are highly incomplete for $\lambda < 20$, we present results only for $\lambda \geq 20$. The relations for these clusters are shown in Figs.~\ref{lx_lambda_full_l20} and~\ref{triangle_l20}. The scatter in the relation is comparable between clusters with and without substructure. We note some differences in the slope and normalization; however, the values compensate for each other, making the relation comparable between the samples. We check for further redshift and richness trends below. Since \citet{Damsted23} report large scatter in scaling relations at $z < 0.2$, we first tested how substructure contributes to these results.

\begin{figure*}
\centering
\begin{tabular}{c}
\resizebox{\textwidth}{!}{\includegraphics{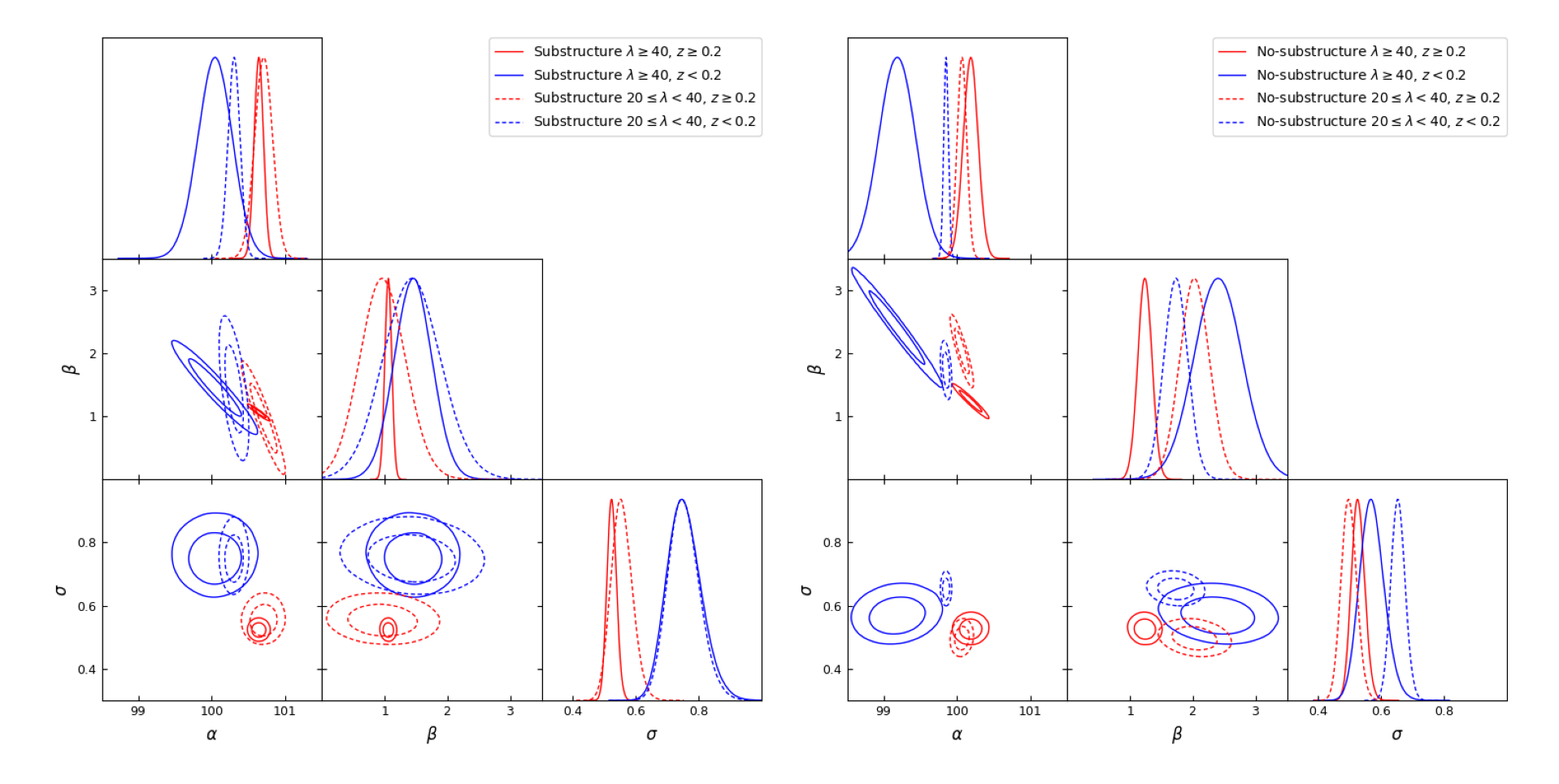}}
\end{tabular}
\caption{Effect of substructure on the redshift and the richness-dependent trends on the X-ray luminosity vs. richness scaling relation. Left panel: Clusters with substructure. Right panel: Clusters without substructure. Blue shows $z < 0.2$ and red $z \geq 0.2$. Solid lines show $\lambda \geq 40$ samples, and dashed lines $20 \leq \lambda < 40$ samples.}
\label{sub_vs_nosub_triangle}
\end{figure*}

Figure~\ref{sub_vs_nosub_triangle} shows redshift-dependent trends for two richness bins, separately for clusters with and without substructure. Clusters without substructure exhibit very mild evolution in scatter with redshift, while those with substructure show strong evolution in both richness bins. We also observe a higher normalization in the scaling relations for clusters with substructure, with the difference being comparable to the scatter in that subsample. Notably, this effect is only pronounced at low redshifts. If interpreted in the context of the growing importance of cool cores at low redshifts \citep{McDonald13}, the excess luminosity in merging clusters could be naturally explained, along with the dominance of cool cores at low $z$. One may further speculate whether survival times for low-entropy components increase at low redshifts, possibly due to stronger magnetic fields, which can be further verified by corresponding predictions for redshift evolution of the nonthermal radio emission from galaxy clusters (see \citealt{vanWeeren19} for a review).

\begin{figure}
\centering
\begin{tabular}{c}
\resizebox{0.45\textwidth}{!}{\includegraphics{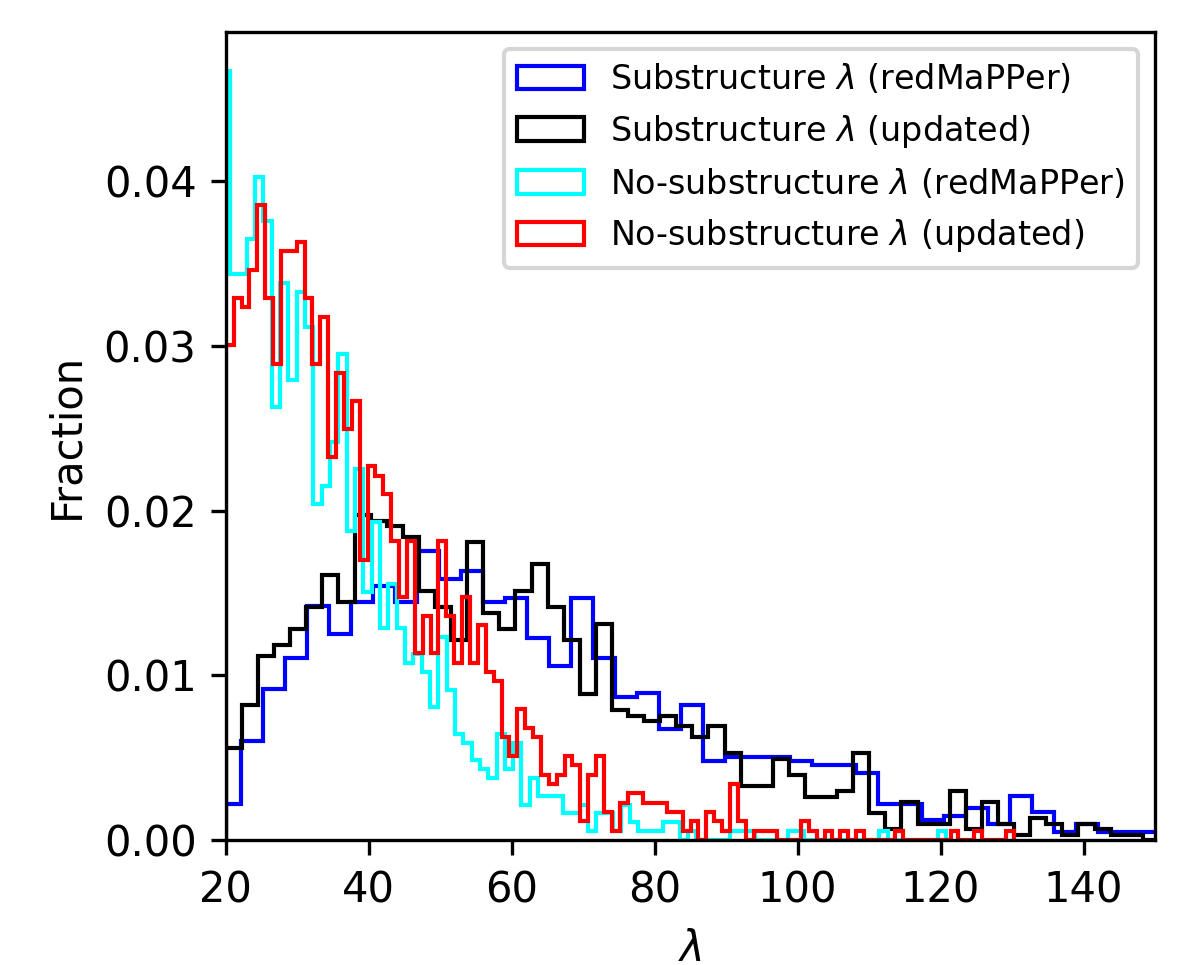}}
\end{tabular}
\caption{Comparison of the redMaPPer and updated $\lambda$ distributions for clusters with and without substructure.}
\label{richness_comparison}
\end{figure}

The richness distributions---including original redMaPPer values and our updated estimates (as described in Sect.~\ref{hdbscan}) for clusters with and without substructure---are shown in Fig.~\ref{richness_comparison}. Clusters with substructure are slightly cleaner after HDBSCAN processing than those without, which partly contributes to the higher normalization but does not fully explain it, especially the redshift dependence observed in Fig.~\ref{sub_vs_nosub_triangle}. Fitting the relation using the original {redMaPPer} richness, we find that using our updated HDBSCAN-based richness reduces the scatter by 7\% for clusters with $\lambda \geq 20$.

\begin{figure}
\centering
\begin{tabular}{c}
\resizebox{0.45\textwidth}{!}{\includegraphics{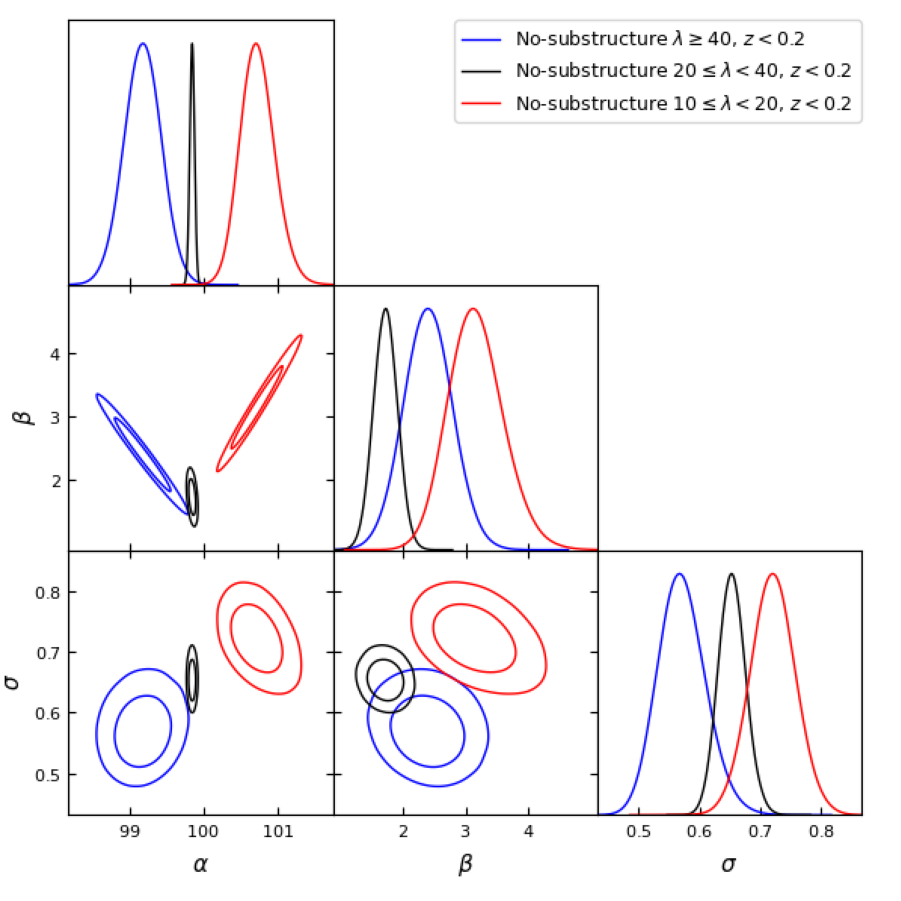}}
\end{tabular}
\caption{Effect of richness on the {X-ray luminosity vs. richness scaling relation. Blue shows the $\lambda \geq 40$ clusters, black $20 \leq \lambda < 40$ clusters, and red $10 \leq \lambda < 20$ clusters. All clusters are without substructure.}}
\label{nosub_triangle_lowz}
\end{figure}

Additionally, we find that scatter is a strong function of richness. This is evident in Fig.~\ref{nosub_triangle_lowz}, which shows no-substructure clusters at $z < 0.2$ in three richness bins, including the range $10 \leq \lambda < 20$ (which is well sampled at low $z$). A clear decrease in scatter with increasing richness is observed. The other parameters show strong correlation and the values are consistent within $2\sigma$. We have marginal evidence that the relation is steeper for low-mass clusters.

\begin{figure*}
\centering
\begin{tabular}{c}
\resizebox{\textwidth}{!}{\includegraphics{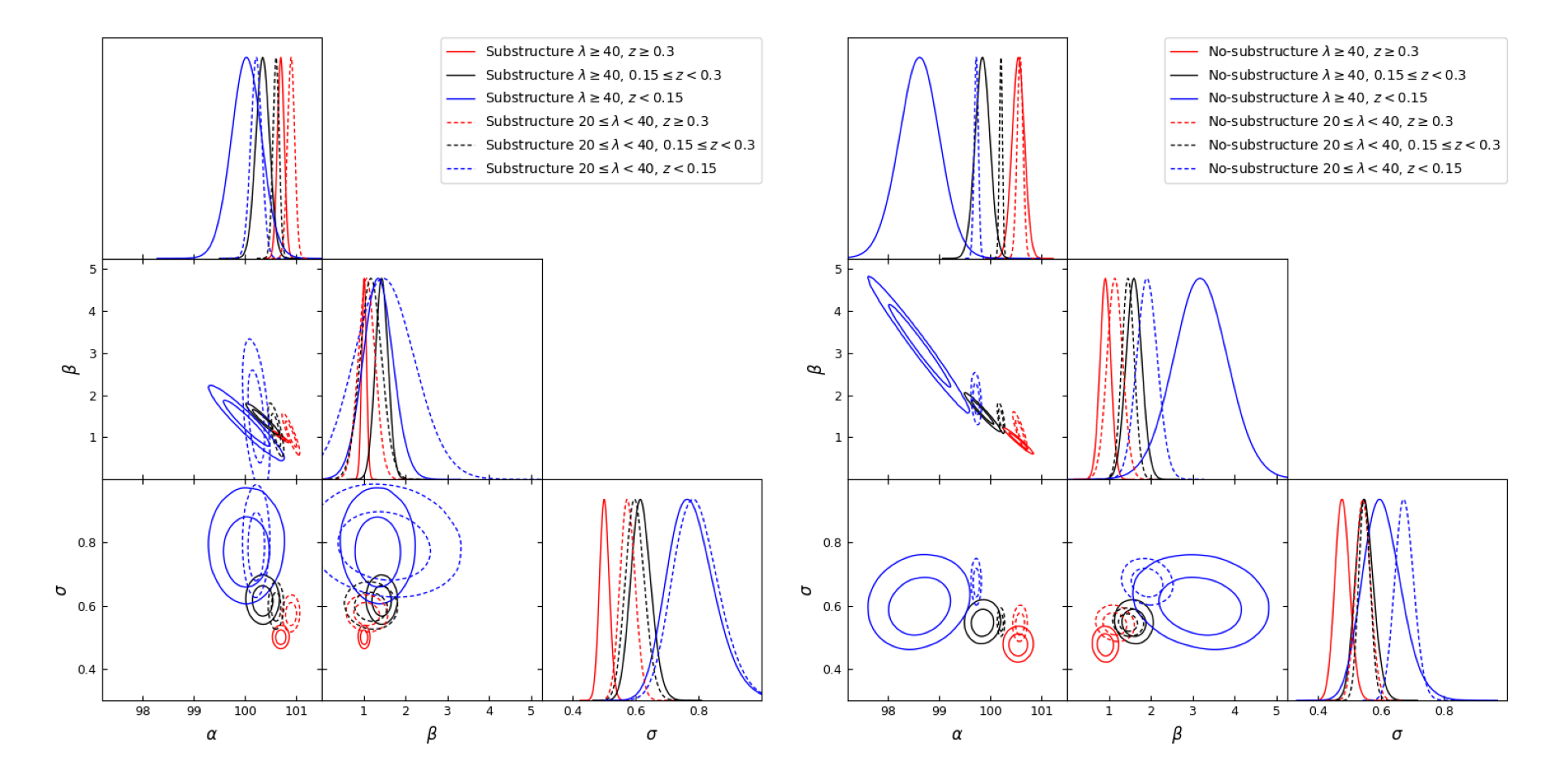}}
\end{tabular}
\caption{Effect of substructure on the redshift and richness-dependent trends on the $L_x-\lambda$ scaling relation without the $\lambda > z\times100$ limitation. Left panel: Clusters with substructure. Right panel: Clusters without substructure. Red shows $z\geq0.3$, black $0.15 \leq z < 0.3,$ and blue $z<0.15$. Solid lines show $\lambda \geq 40$ samples, and dashed lines $20\leq\lambda < 40$ samples.}
\label{sub_vs_nosub_triangle_no_cut}
\end{figure*}

Our results are not affected by the limitation $\lambda > z\times100$. We can observe the same redshift and richness-dependent trends in Fig.~\ref{sub_vs_nosub_triangle_no_cut} where the limitation was not applied.

We also fit the scaling relation to the subsamples of multiple associated sources described in Sect.~\ref{mult}. For the subsample of multiple X-ray sources (i.e., optical clusters matched to more than one eROSITA detection), we included only those systems where the substructure identified by HDBSCAN aligns well with the multiple X-ray sources and the $\lambda > z\times100$ condition is fulfilled. This left us with 41 subclusters. Of these, 34 are at $z < 0.2$. The mean subcluster richness of this sample is $\bar{\lambda} = 26.3$. Here, the subcluster richness is not limited to $\lambda \geq 20$. We find that these clusters exhibit a relation comparable to that of the general substructure sample with similar richness and redshift.

For the subsample of multiple optical clusters (i.e., X-ray sources matched to more than one optical cluster), we summed the richness values ($\lambda$) of all associated optical clusters within a single X-ray source to form a total richness estimate. This combined richness is used in the scaling relation fit. The mean richness and redshift of the subsample are $\bar{\lambda}=71.4$ and $\bar{z}=0.31$. We find that this subsample exhibits a relation comparable to the general substructure sample of massive clusters ($\lambda \geq 40$), with slightly increased scatter for a given redshift.

\begin{table*}
    \caption{Regression analysis for X-ray luminosity vs. richness scaling relation $\ln(L_xE_z^{-1} \, \mathrm{erg^{-1} \, s}) = \alpha + \beta\ln(\lambda/27.9 \cdot E_z) + N(0, \sigma)$.}
    \label{tab:scaling_relations}
    \begin{tabular}{l c c c c}
        \hline\hline
         Sample & Intercept $\alpha$ & Slope $\beta$ & Intrinsic scatter $\sigma$ & N clusters \\
         \hline
         $\lambda \geq 20$ {(updated $\lambda$ estimates)} & 100.094 $\pm$ 0.020 & 1.463 $\pm$ 0.025 & 0.613 $\pm$ 0.009 & 2951 \\
         $\lambda \geq 20$ (redMaPPer $\lambda$) & 100.403 $\pm$ 0.017 & 1.140 $\pm$ 0.022 & 0.659 $\pm$ 0.010 & 2951 \\
         Substructure $\lambda \geq 20$ & 100.492 $\pm$ 0.035 & 1.168 $\pm$ 0.036 & 0.586 $\pm$ 0.013 & 1352 \\
         No-substructure $\lambda \geq 20$ & 99.957 $\pm$ 0.024 & 1.497 $\pm$ 0.042 & 0.592 $\pm$ 0.013 & 1599 \\
         Substructure $\lambda \geq 40$ & 100.497 $\pm$ 0.060 & 1.160 $\pm$ 0.054 & 0.564 $\pm$ 0.015 & 1015 \\
         No-substructure $\lambda \geq 40$ & 99.899 $\pm$ 0.086 & 1.505 $\pm$ 0.102 & 0.555 $\pm$ 0.019 & 616 \\
         Substructure $20 \leq \lambda < 40$ & 100.399 $\pm$ 0.066 & 1.660 $\pm$ 0.240 & 0.653 $\pm$ 0.030 & 337 \\
         No-substructure $20 \leq \lambda < 40$ & 99.904 $\pm$ 0.028 & 2.134 $\pm$ 0.132 & 0.609 $\pm$ 0.017 & 983 \\ [1.0ex]

         Substructure $\lambda \geq 40$, $z \geq 0.2$ & 100.642 $\pm$ 0.064 & 1.061 $\pm$ 0.055 & 0.523 $\pm$ 0.015 & 897 \\
         Substructure $\lambda \geq 40$, $z < 0.2$ & 100.046 $\pm$ 0.238 & 1.457 $\pm$ 0.303 & 0.752 $\pm$ 0.056 & 118 \\
         No-substructure $\lambda \geq 40$, $z \geq 0.2$ & 100.186 $\pm$ 0.101 & 1.239 $\pm$ 0.112 & 0.526 $\pm$ 0.021 & 463 \\
         No-substructure $\lambda \geq 40$, $z < 0.2$ & 99.181 $\pm$ 0.253 & 2.406 $\pm$ 0.387 & 0.570 $\pm$ 0.040 & 153 \\ [1.0ex]

         Substructure $20 \leq \lambda < 40$, $z \geq 0.2$ & 100.707 $\pm$ 0.122 & 0.969 $\pm$ 0.367 & 0.554 $\pm$ 0.034 & 195 \\
         Substructure $20 \leq \lambda < 40$, $z < 0.2$ & 100.305 $\pm$ 0.083 & 1.436 $\pm$ 0.464 & 0.751 $\pm$ 0.051 & 142 \\
         No-substructure $20 \leq \lambda < 40$, $z \geq 0.2$ & 100.066 $\pm$ 0.066 & 2.031 $\pm$ 0.237 & 0.497 $\pm$ 0.025 & 396 \\
         No-substructure $20 \leq \lambda < 40$, $z < 0.2$ & 99.850 $\pm$ 0.032 & 1.733 $\pm$ 0.190 & 0.653 $\pm$ 0.023 & 587 \\
         No-substructure $10 \leq \lambda < 20$, $z < 0.2$ & 100.727 $\pm$ 0.234 & 3.149 $\pm$ 0.440 & 0.721 $\pm$ 0.037 & 396 \\
         [1.0ex]

         Multiple X-rays subsample & 100.457 $\pm$ 0.154 & 1.277 $\pm$ 0.333 & 0.905 $\pm$ 0.127 & 41 \\
         Multiple opticals subsample ($\lambda \geq 20$)& 100.432 $\pm$ 0.097 & 1.181 $\pm$ 0.087 & 0.709 $\pm$ 0.033 & 270 \\

         \hline        
    \end{tabular}
    \tablefoot{Uncertainties are quoted for the $68\%$ confidence interval
    }
\end{table*}

\begin{table*}
    \caption{Main results from the $L_x-\lambda$ scaling relation analysis.}
    \label{tab:scaling_summary}
    \begin{tabular}{l c c l c}
        \hline\hline
        Test sample & Result & {Probability} & Baseline sample & Parameter values  \\
        \hline
        Sub $\lambda \geq 40$, $z < 0.2$ & Higher $\alpha$ & 0.994 & No-sub $\lambda \geq 40$, $z < 0.2$ & 100.046 $\pm$ 0.238 vs. 99.181 $\pm$ 0.253 \\
        Sub $20 \leq \lambda < 40$, $z < 0.2$ & Higher $\alpha$ & 1.000 & No-sub $20 \leq \lambda < 40$, $z < 0.2$ & 100.305 $\pm$ 0.083 vs. 99.850 $\pm$ 0.032 \\ [0.5ex]
         
         Sub $\lambda \geq 40$, $z < 0.2$ & Higher $\sigma$ & 0.998 & No-sub $\lambda \geq 40$, $z < 0.2$ & 0.752 $\pm$ 0.056 vs. 0.570 $\pm$ 0.040 \\
        Sub $20 \leq \lambda < 40$, $z < 0.2$ & Higher $\sigma$ & 0.971 & No-sub $20 \leq \lambda < 40$, $z < 0.2$ & 0.751 $\pm$ 0.051 vs. 0.653 $\pm$ 0.023 \\ [0.5ex]

         $\lambda \geq 20$ (updated $\lambda$) & Lower $\sigma$ & 0.999 & $\lambda \geq 20$  (redMaPPer $\lambda$) & 0.613 $\pm $0.009 vs. 0.659 $\pm$ 0.010 \\
         \hline
    \end{tabular}
    \tablefoot{Here ``sub'' refers to clusters with detected substructure and ``No-sub'' to clusters without substructure. The probability column shows the probability that the test sample parameter is higher or lower, as denoted by the second column, than the baseline sample.}
\end{table*}

\section{Conclusions}\label{conclusion}

Based on our comprehensive analysis of substructure in galaxy clusters using HDBSCAN on the {redMaPPer} catalog and its correlation with X-ray morphology and scaling relations, we draw the following conclusions:
\begin{itemize}
\item
Our two-step HDBSCAN procedure successfully identifies substructure in a significant fraction of clusters, revealing 2159 clusters with substructure (containing between 2 and 11 subclusters) and 4262 without. The method effectively cleans galaxy membership assignments, reducing scatter in richness estimates. The mass ratio $\mu = M_{\text{infalling}}/M_{\text{host}}$ is a crucial parameter, with completeness limits of $\mu > 0.4$ for $\lambda \geq 20$ and $\mu > 0.25$ for $\lambda \geq 40$.

\item Our substructure detections show considerable agreement with independent catalogs of merging clusters \citep{Wen2024}, validating our methodology. Differences arise primarily from the initial cluster membership definition in {redMaPPer}, which is limited to the virial radius ($R_{200c}$) and may filter out very early-stage mergers. Thus, our results characterize substructure within the virialized region of clusters.

\item
Clusters with substructure are unequivocally more disturbed. They exhibit significantly different distributions in key morphological parameters, particularly flatter surface brightness profiles (lower concentration $c_{80-800}$), higher centroid shifts ($\omega$), and larger multipole moments ($M_4$). The combined disturbance score ($D_{\text{comb}}$), which incorporates concentration, shows a far stronger distinction than the shape-based $D_{\text{shape}}$ score. This confirms that substructure is a primary driver of morphological disturbance in clusters.
\end{itemize}

\noindent The presence of substructure significantly affects the $L_X-\lambda$ scaling relation. We find:
\begin{itemize}

\item Clusters with substructure exhibit a higher normalization in the $L_X-\lambda$ relation at low redshifts ($z < 0.2$). This is consistent with the survival of cool cores from infalling subclusters, which boost the X-ray luminosity, and aligns with the known increased prevalence of cool cores at low redshifts.

\item Substructure introduces greater intrinsic scatter into the scaling relation at low redshifts. The survival of cool cores from infalling subclusters produces highly luminous systems, where at times multiple cool cores can even be present. Occasionally, the cool cores are destroyed in the mergers, which leads to under-luminous systems. The presence of both over- and under-luminous systems increases the scatter.

\item The effect of substructure on scaling relations is strongly redshift-dependent. The increased scatter and normalization are dominant at low redshifts, while at higher redshifts the differences between clusters with and without substructure become less distinct, likely due to the declining prevalence of stable cool cores.

\item The ability to detect substructure and its impact is a strong function of both cluster richness and redshift. Substructure is more common in high-richness systems. Furthermore, morphological parameters themselves correlate with richness; lower-richness clusters are generally less spherical and symmetrical regardless of substructure.
\end{itemize}

In summary, internal substructure is a fundamental property of galaxy clusters that must be accounted for in cosmological studies. It is a major source of scatter in scaling relations and a significant contributor to morphological disturbance, particularly at low redshifts. Correcting for its effects, through methods like the one presented here, is essential for achieving the precision required in next-generation cluster cosmology. Future efforts incorporating spectroscopic data for 3D clustering and larger samples from upcoming surveys will further refine our understanding of substructure and its role in cluster evolution.

\section{Data Availability}

The catalogs of the identified redMaPPer substructure are available at the CDS via anonymous ftp to \url{cdsarc.cds.unistra.fr}.

   \bibliographystyle{aa} 
   \bibliography{ref} 

\begin{appendix}
\section{Acknowledgements}
\begin{acknowledgements}

We thank the referee for the insightful comments, which helped us to improve this paper. 
The work is based on the Legacy Survey data.
The Legacy Surveys consist of three individual and complementary projects: the Dark Energy Camera Legacy Survey (DECaLS; Proposal ID \#2014B-0404; PIs: David Schlegel and Arjun Dey), the Beijing-Arizona Sky Survey (BASS; NOAO Prop. ID \#2015A-0801; PIs: Zhou Xu and Xiaohui Fan), and the Mayall z-band Legacy Survey (MzLS; Prop. ID \#2016A-0453; PI: Arjun Dey). DECaLS, BASS and MzLS together include data obtained, respectively, at the Blanco telescope, Cerro Tololo Inter-American Observatory, NSF’s NOIRLab; the Bok telescope, Steward Observatory, University of Arizona; and the Mayall telescope, Kitt Peak National Observatory, NOIRLab. Pipeline processing and analyses of the data were supported by NOIRLab and the Lawrence Berkeley National Laboratory (LBNL). The Legacy Surveys project is honored to be permitted to conduct astronomical research on Iolkam Du’ag (Kitt Peak), a mountain with particular significance to the Tohono O’odham Nation.

NOIRLab is operated by the Association of Universities for Research in Astronomy (AURA) under a cooperative agreement with the National Science Foundation. LBNL is managed by the Regents of the University of California under contract to the U.S. Department of Energy.

This project used data obtained with the Dark Energy Camera (DECam), which was constructed by the Dark Energy Survey (DES) collaboration. Funding for the DES Projects has been provided by the U.S. Department of Energy, the U.S. National Science Foundation, the Ministry of Science and Education of Spain, the Science and Technology Facilities Council of the United Kingdom, the Higher Education Funding Council for England, the National Center for Supercomputing Applications at the University of Illinois at Urbana-Champaign, the Kavli Institute of Cosmological Physics at the University of Chicago, Center for Cosmology and Astro-Particle Physics at the Ohio State University, the Mitchell Institute for Fundamental Physics and Astronomy at Texas A\&M University, Financiadora de Estudos e Projetos, Fundacao Carlos Chagas Filho de Amparo, Financiadora de Estudos e Projetos, Fundacao Carlos Chagas Filho de Amparo a Pesquisa do Estado do Rio de Janeiro, Conselho Nacional de Desenvolvimento Cientifico e Tecnologico and the Ministerio da Ciencia, Tecnologia e Inovacao, the Deutsche Forschungsgemeinschaft and the Collaborating Institutions in the Dark Energy Survey. The Collaborating Institutions are Argonne National Laboratory, the University of California at Santa Cruz, the University of Cambridge, Centro de Investigaciones Energeticas, Medioambientales y Tecnologicas-Madrid, the University of Chicago, University College London, the DES-Brazil Consortium, the University of Edinburgh, the Eidgenossische Technische Hochschule (ETH) Zurich, Fermi National Accelerator Laboratory, the University of Illinois at Urbana-Champaign, the Institut de Ciencies de l’Espai (IEEC/CSIC), the Institut de Fisica d’Altes Energies, Lawrence Berkeley National Laboratory, the Ludwig Maximilians Universitat Munchen and the associated Excellence Cluster Universe, the University of Michigan, NSF’s NOIRLab, the University of Nottingham, the Ohio State University, the University of Pennsylvania, the University of Portsmouth, SLAC National Accelerator Laboratory, Stanford University, the University of Sussex, and Texas A\&M University.

BASS is a key project of the Telescope Access Program (TAP), which has been funded by the National Astronomical Observatories of China, the Chinese Academy of Sciences (the Strategic Priority Research Program “The Emergence of Cosmological Structures” Grant \# XDB09000000), and the Special Fund for Astronomy from the Ministry of Finance. The BASS is also supported by the External Cooperation Program of Chinese Academy of Sciences (Grant \# 114A11KYSB20160057), and Chinese National Natural Science Foundation (Grant \# 12120101003, \# 11433005).

The Legacy Survey team makes use of data products from the Near-Earth Object Wide-field Infrared Survey Explorer (NEOWISE), which is a project of the Jet Propulsion Laboratory/California Institute of Technology. NEOWISE is funded by the National Aeronautics and Space Administration.

The Legacy Surveys imaging of the DESI footprint is supported by the Director, Office of Science, Office of High Energy Physics of the U.S. Department of Energy under Contract No. DE-AC02-05CH1123, by the National Energy Research Scientific Computing Center, a DOE Office of Science User Facility under the same contract; and by the U.S. National Science Foundation, Division of Astronomical Sciences under Contract No. AST-0950945 to NOAO.

This work is based on data from eROSITA, the soft X-ray instrument aboard SRG, a joint Russian-German science mission supported by the Russian Space Agency (Roskosmos), in the interests of the Russian Academy of Sciences represented by its Space Research Institute (IKI), and the Deutsches Zentrum für Luft- und Raumfahrt (DLR). The SRG spacecraft was built by Lavochkin Association (NPOL) and its subcontractors, and is operated by NPOL with support from the Max Planck Institute for Extraterrestrial Physics (MPE). The development and construction of the eROSITA X-ray instrument was led by MPE, with contributions from the Dr. Karl Remeis Observatory Bamberg \& ECAP (FAU Erlangen-Nuernberg), the University of Hamburg Observatory, the Leibniz Institute for Astrophysics Potsdam (AIP), and the Institute for Astronomy and Astrophysics of the University of Tübingen, with the support of DLR and the Max Planck Society. The Argelander Institute for Astronomy of the University of Bonn and the Ludwig Maximilians Universität München also participated in the science preparation for eROSITA.
\end{acknowledgements}
\section{Tables}
\begin{figure}[p!]
\begin{tabular}{c}
\resizebox{\hsize}{!}{\includegraphics{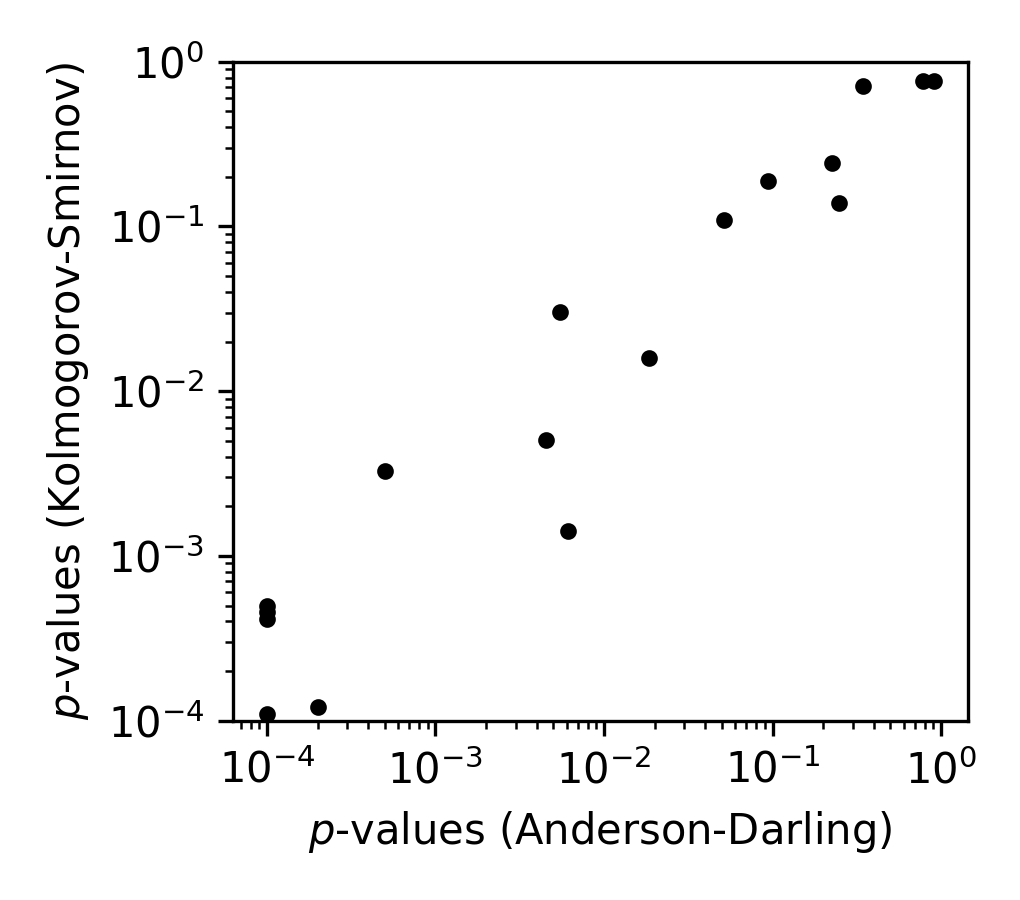}}
\end{tabular}
\caption{$p$-values from the AD and KS tests plotted against each other.}
\label{ad_vs_ks}
\end{figure}

Table \ref{parameters_description} summarizes all of the morphological parameters from \citet{Sanders2025}. Tables \ref{p_values_full}-\ref{p_values_l40} contain all of the $p$-values from the performed statistical tests. The statistical tests were performed using the SciPy \citep{2020SciPy-NMeth} modules for Python. Due to the finite number of permutations, the $p$-values provided by the AD and permutation tests are capped at $1.0\times10^{-4}$, while the KS-test has no such limitation. Figure \ref{ad_vs_ks} shows the $p$-values from the AD and KS tests plotted against each other. Here entries with KS $p$-values lower than the AD-test cap of $1.0\times10^{-4}$ are left out in order to make a reasonable comparison. We find that the tests are generally in good agreement.
\begin{table*}[h!]
\caption{Morphological parameters adopted from \citet{Sanders2025}.}
\label{parameters_description}
\begin{tabular}{l l}
\hline
Name &  Description \\
\hline
$n_{s,0}$ & Log gas density at a radius of $0.02R_{500}$ relative to scaled critical density \\
$n_{s,0}^*$ & Log gas density at a radius of $0.02R_{500}$ relative to scaled critical density \\
$n_{50}$ & Log gas electron density at a radius of 50 kpc \\
$n_{50}^*$ & Log gas electron density at a radius of 50 kpc \\
$\alpha$ & Inner density slope at $0.04R_{500}$ radius \\
$\alpha^*$ & Inner density slope at $0.04R_{500}$ radius \\
$\alpha_{50}$ & Inner density slope at 50 kpc radius \\
$\alpha_{50}^*$ & Inner density slope at 50 kpc radius \\
$c_{500}$ & Log ratio of integrated model surface brightness in apertures of $0.1R_{500}$ and $R_{500}$ \\
$c_{500}^*$ & Log ratio of integrated model surface brightness in apertures of $0.1R_{500}$ and $R_{500}$ \\
$c_{80-800}$ & Log ratio of integrated model surface brightness in apertures of 80 and 800 kpc \\
$c_{80-800}^*$ & Log ratio of integrated model surface brightness in apertures of 80 and 800 kpc \\
$F$ & Ratio of offset between best fitting cluster center and peak and $R_{500}$ \\
$P_{10}$, $P_{20}$, $P_{30}$, $P_{40}$ & Log power ratio with orders from 1 to 4 \\
$P_{10}^*$, $P_{20}^*$ & Log power ratio with orders from 1 to 2 \\
$G$ & Gini coefficient (0-1) \\
$A_\text{phot}$ & Photon asymmetry \\
$A_\text{phot}^*$ & Photon asymmetry \\
$\omega$ & Centroid shift \\
$\epsilon$ & Ellipticity, the ratio of minor to major axis (0-1) \\
$H$ & Slosh, the degree of sloshing factor (0-1) \\
$M_1, M_2, M_3, M_4$ & Multipole magnitudes (0-1) for orders 1 to 4 \\
\hline
$D_\text{Shape}$ & Combined disturbance score (0-1), based on $\epsilon$, $H$, $M_1$ to $M_4$ and $F$ \\
$D_\text{Comb}$ & Combined disturbance score (0-1), also including $c_{80-800}$ \\
\hline
\end{tabular}
\end{table*}

\FloatBarrier

\twocolumn
\begin{table}[p!]
\caption{$p$-values from the {AD, KS,} and shuffle tests {comparing} the substructure and no-substructure samples {for the full $\lambda \geq 20$ sample without any $\mu$ cuts}.}
\label{p_values_full}
\resizebox{0.45\textwidth}{!}{\begin{tabular}{l c c c c}
\hline
Name & {AD} & KS & Shuffle (mean) & Shuffle (median) \\
\hline
$n_{s,0}$ & 0.018 & 0.016 & 0.041 & 0.081 \\
$n_{s,0}^*$ & $5.00\times10^{-3}$ & $5.06\times10^{-3}$ & 0.36 & 0.38 \\
$n_{50}$ & $1.00\times10^{-4}$ & $2.28\times10^{-28}$ & $2.00\times10^{-4}$ & $2.00\times10^{-4}$  \\
$n_{50}^*$ & $1.00\times10^{-4}$ & $4.05\times10^{-50}$ & $2.00\times10^{-4}$ & $2.00\times10^{-4}$ \\
$\alpha$ & $1.00\times10^{-4}$ & $2.11\times10^{-6}$ & $2.00\times10^{-4}$ & $2.00\times10^{-4}$ \\
$\alpha^*$ & $2.00\times10^{-4}$ & $1.95\times10^{-9}$ & $2.00\times10^{-4}$ & $2.00\times10^{-4}$ \\
$\alpha_{50}$ & $1.00\times10^{-4}$ & $3.40\times10^{-14}$ & $2.00\times10^{-4}$ & $2.00\times10^{-4}$ \\
$\alpha_{50}^*$ & $1.00\times10^{-4}$ & $2.75\times10^{-18}$ & $2.00\times10^{-4}$ & $2.00\times10^{-4}$\\
$c_{500}$ & $1.00\times10^{-4}$ & $4.95\times10^{-4}$ & $2.00\times10^{-4}$ & $2.00\times10^{-4}$ \\
$c_{500}^*$ & $1.00\times10^{-4}$ & $1.10\times10^{-4}$ & $2.00\times10^{-4}$ & $2.00\times10^{-4}$ \\
$c_{80-800}$ & $1.00\times10^{-4}$ & $3.22\times10^{-17}$ & $2.00\times10^{-4}$ & $2.00\times10^{-4}$ \\
$c_{80-800}^*$ & $1.00\times10^{-4}$ & $2.16\times10^{-14}$ & $2.00\times10^{-4}$ & $2.00\times10^{-4}$ \\
$F$ & $1.00\times10^{-4}$ & $4.12\times10^{-4}$ & 0.62 & $6.00\times10^{-4}$ \\
$P_{10}$ & $1.00\times10^{-4}$ & $1.39\times10^{-5}$ & $2.00\times10^{-4}$ & $1.00\times10^{-3}$ \\
$P_{20}$ & $1.00\times10^{-4}$ & $8.60\times10^{-5}$ & $2.00\times10^{-4}$ & $6.00\times10^{-4}$ \\
$P_{30}$ & $1.00\times10^{-4}$ & $2.52\times10^{-5}$ & $2.00\times10^{-4}$ & $2.00\times10^{-4}$ \\
$P_{40}$ & $1.00\times10^{-4}$ & $1.12\times10^{-5}$ & $2.00\times10^{-4}$ & $2.00\times10^{-4}$ \\
$P_{10}^*$ & $4.80\times10^{-3}$ & 0.030 & $1.40\times10^{-3}$ & $6.00\times10^{-3}$ \\
$P_{20}^*$ & $1.00\times10^{-4}$ & $9.25\times10^{-5}$ & $2.00\times10^{-4}$ & $1.40\times10^{-3}$ \\
$G$ & $1.00\times10^{-4}$ & $1.21\times10^{-4}$ & $2.60\times10^{-3}$ & $1.00\times10^{-3}$\\
$A_\text{phot}$ & $1.00\times10^{-4}$ & $6.21\times10^{-6}$ & 0.60 & $2.00\times10^{-4}$ \\
$A_\text{phot}^*$ & $1.00\times10^{-4}$ & $4.59\times10^{-4}$ & 0.457 & $4.00\times10^{-4}$ \\
$\omega$ & $4.00\times10^{-4}$ & $3.27\times10^{-3}$ & $8.00\times10^{-4}$ & 0.021 \\
$\epsilon$ & $5.30\times10^{-3}$ & $1.41\times10^{-3}$ & $7.80\times10^{-3}$ & $7.80\times10^{-3}$ \\
$H$ & 0.91 & 0.77 & 0.61 & 0.40 \\
$M_1$ & 0.34 & 0.71 & 0.47 & 0.69 \\
$M_2$ & 0.24 & 0.14 & 0.15 & 0.047 \\
$M_3$ & 0.78 & 0.76 & 0.38 & 0.36 \\
$M_4$ & 0.22 & 0.24 & 0.60 & 0.13 \\
\hline
$D_\text{shape}$ & 0.049 & 0.11 & 0.027 & 0.10 \\
$D_\text{comb}$ & $1.00\times10^{-4}$ & $5.59\times10^{-12}$ & $2.00\times10^{-4}$ & $2.00\times10^{-4}$ \\
\hline
\end{tabular}}
\end{table}

\begin{table}[p!]
\caption{{$p$-values from the AD, KS, and shuffle tests {comparing} the substructure and no-substructure samples for the $20 \leq \lambda < 40$ richness bin.}}
\label{p_values_l2040}
\resizebox{0.45\textwidth}{!}{\begin{tabular}{l c c c c}
\hline
Name & AD & KS & Shuffle (mean) & Shuffle (median) \\
\hline
$n_{s,0}$ & 0.053 & 0.030 & 0.19 & 0.023 \\
$n_{s,0}^*$ & 0.027 & 0.071 & 0.56 & 0.092 \\
$n_{50}$ & $1.00\times10^{-4}$ & $7.73\times10^{-10}$ & $2.00\times10^{-4}$ & $2.00\times10^{-4}$ \\
$n_{50}^*$ & $1.00\times10^{-4}$ & $7.39\times10^{-18}$ & $2.00\times10^{-4}$ & $2.00\times10^{-4}$ \\
$\alpha$ & $1.10\times10^{-3}$ & $2.33\times10^{-3}$ & $8.00\times10^{-4}$ & $8.00\times10^{-4}$ \\
$\alpha^*$ & $1.00\times10^{-4}$ & $6.82\times10^{-5}$ & $6.00\times10^{-4}$ & $2.00\times10^{-4}$ \\
$\alpha_{50}$ & $1.00\times10^{-4}$ & $2.21\times10^{-6}$ & $2.00\times10^{-4}$ & $2.00\times10^{-4}$ \\
$\alpha_{50}^*$ & $1.00\times10^{-4}$ & $1.06\times10^{-8}$ & $2.00\times10^{-4}$ & $2.00\times10^{-4}$ \\
$c_{500}$ & $2.40\times10^{-3}$ & $3.63\times10^{-3}$ & $3.80\times10^{-3}$ & $2.60\times10^{-3}$ \\
$c_{500}^*$ & $3.10\times10^{-3}$ & $8.40\times10^{-3}$ & $6.20\times10^{-3}$ & $3.60\times10^{-3}$ \\
$c_{80-800}$ & $1.00\times10^{-4}$ & $2.44\times10^{-7}$ & $2.00\times10^{-4}$ & $2.00\times10^{-4}$ \\
$c_{80-800}^*$ & $1.00\times10^{-4}$ & $3.26\times10^{-5}$ & $2.00\times10^{-4}$ & $2.00\times10^{-4}$ \\
$F$ & $8.00\times10^{-3}$ & $1.60\times10^{-3}$ & 0.72 & $8.00\times10^{-4}$ \\
$P_{10}$ & 0.11 & 0.059 & 0.044 & 0.23 \\
$P_{20}$ & 0.031 & 0.038 & 0.039 & 0.16 \\
$P_{30}$ & 0.039 & 0.073 & 0.027 & 0.25 \\
$P_{40}$ & 0.10 & 0.24 & 0.044 & 0.12 \\
$P_{10}^*$ & 0.46 & 0.30 & 0.38 & 0.60 \\
$P_{20}^*$ & 0.031 & $5.77\times10^{-3}$ & 0.077 & 0.16 \\
$G$ & 0.081 & 0.18 & 0.29 & 0.32 \\
$A_\text{phot}$ & $3.60\times10^{-3}$ & 0.017 & 0.85 & 0.060 \\
$A_\text{phot}^*$ & $7.40\times10^{-3}$ & 0.019 & 0.95 & 0.28 \\
$\omega$ & $2.40\times10^{-3}$ & 0.016 & $2.40\times10^{-3}$ & 0.019 \\
$\epsilon$ & 0.14 & 0.08 & 0.31 & 0.15 \\
$H$ & 0.90 & 0.71 & 0.87 & 0.25 \\
$M_1$ & 0.89 & 0.93 & 0.63 & 0.73 \\
$M_2$ & 0.19 & 0.47 & 0.75 & 0.47 \\
$M_3$ & 0.92 & 0.96 & 0.82 & 0.68 \\
$M_4$ & 0.44 & 0.27 & 0.62 & 0.19 \\
\hline
$D_\text{shape}$ & 0.60 & 0.42 & 0.46 & 0.86 \\
$D_\text{comb}$ & $1.00\times10^{-4}$ & $9.95\times10^{-7}$ & $2.00\times10^{-4}$ & $2.00\times10^{-4}$ \\
\hline
\end{tabular}}
\end{table}

\begin{table}[p!]
\caption{{$p$-values from the AD, KS, and shuffle tests {comparing} the substructure and no-substructure samples for the $\lambda \geq 40$ richness bin.}}
\label{p_values_l40}
\resizebox{0.45\textwidth}{!}{\begin{tabular}{l c c c c}
\hline
Name & AD & KS & Shuffle (mean) & Shuffle (median) \\
\hline
$n_{s,0}$ & 0.21 & 0.47 & 0.54 & 0.85 \\
$n_{s,0}^*$ & 0.13 & 0.28 & 0.81 & 0.54 \\
$n_{50}$ & $1.00\times10^{-4}$ & $1.28\times10^{-6}$ & $8.00\times10^{-4}$ & $2.00\times10^{-4}$ \\
$n_{50}^*$ & $1.00\times10^{-4}$ & $6.30\times10^{-10}$ & $2.00\times10^{-4}$ & $2.00\times10^{-4}$ \\
$\alpha$ & 0.042 & 0.044 & 0.036 & 0.50 \\
$\alpha^*$ & $9.80\times10^{-3}$ & $8.58\times10^{-3}$ & $8.00\times10^{-3}$ & 0.078 \\
$\alpha_{50}$ & $2.40\times10^{-3}$ & 0.010 & $1.60\times10^{-3}$ & 0.014 \\
$\alpha_{50}^*$ & $4.00\times10^{-4}$ & $9.52\times10^{-4}$ & $4.00\times10^{-4}$ & $6.40\times10^{-3}$ \\
$c_{500}$ & 0.19 & 0.24 & 0.11 & 0.97 \\
$c_{500}^*$ & 0.21 & 0.37 & 0.18 & 0.43 \\
$c_{80-800}$ & $1.00\times10^{-4}$ & $1.43\times10^{-4}$ & $2.00\times10^{-4}$ & 0.017 \\
$c_{80-800}^*$ & $2.00\times10^{-4}$ & $1.16\times10^{-3}$ & $8.00\times10^{-4}$ & $7.20\times10^{-3}$ \\
$F$ & 0.020 & 0.12 & 0.39 & 0.38 \\
$P_{10}$ & 0.45 & 0.57 & 0.29 & 0.99 \\
$P_{20}$ & 0.48 & 0.75 & 0.40 & 0.81 \\
$P_{30}$ & 0.18 & 0.86 & 0.33 & 0.50 \\
$P_{40}$ & 0.28 & 0.56 & 0.12 & 0.28 \\
$P_{10}^*$ & 0.87 & 0.92 & 0.86 & 0.90 \\
$P_{20}^*$ & 0.49 & 0.86 & 0.33 & 0.50 \\
$G$ & 0.51 & 0.51 & 0.99 & 0.61 \\
$A_\text{phot}$ & 0.32 & 0.33 & 0.78 & 0.38 \\
$A_\text{phot}^*$ & 0.22 & 0.29 & 0.23 & 0.48 \\
$\omega$ & 0.052 & 0.076 & 0.035 & 0.28 \\
$\epsilon$ & 0.74 & 0.45 & 0.56 & 0.81 \\
$H$ & 0.11 & 0.13 & 0.092 & 0.20 \\
$M_1$ & 0.63 & 0.87 & 0.49 & 0.60 \\
$M_2$ & 0.72 & 0.52 & 0.44 & 0.80 \\
$M_3$ & 0.42 & 0.59 & 0.28 & 0.29 \\
$M_4$ & 0.035 & 0.015 & 0.073 & $3.80\times10^{-3}$\\
\hline
$D_\text{shape}$ & 0.016 & 0.019 & 0.021 & 0.012 \\
$D_\text{comb}$ & $9.00\times10^{-4}$ & 0.014 & $4.00\times10^{-4}$ & 0.013 \\
\hline
\end{tabular}}
\end{table}

\end{appendix}
\end{document}